\documentclass[twocolumn,english,prl]{revtex4}
\usepackage[T1]{fontenc}
\usepackage[latin9]{inputenc}
\usepackage{color}
\usepackage{amsmath}
\usepackage{amssymb}
\usepackage{graphicx}
\usepackage{esint}
\usepackage[caption=false]{subfig}

\makeatletter

\newcommand{\ini}{\ensuremath{|\sbb_\text{ini}\rangle}}
\newcommand{\fin}{\ensuremath{|\sbb_\text{fin}\rangle}}
\newcommand{\finT}{\ensuremath{\langle\sbb_\text{fin}|}}

\newcommand{\Ab}{\textbf{A} }		
\newcommand{\Bb}{\textbf{B} }
\newcommand{\Eb}{\textbf{E} }
\newcommand{\Fb}{\textbf{F} }

\newcommand{\ab}{\textbf{a} }

\newcommand{\pb}{\textbf{p} }
\newcommand{\rb}{\textbf{r} }

\newcommand{\vb}{\textbf{v}}

\newcommand{\sbb}{\textbf{s}}

\newcommand{\nb}{\textbf{n}}
\newcommand{\Pb}{\textbf{P}}
\newcommand{\bb}{\textbf{b}}
\newcommand{\tb}{\textbf{t}}

\makeatother

\usepackage{babel}
\begin{document}
	
	\title{Scaling laws for the (de-)polarization time of relativistic particle beams in strong fields}
	
	\author{Johannes Thomas$^{1}$, Anna H\"utzen$^{2,3}$, Andreas Lehrach$^{4,5}$, Alexander Pukhov$^{1}$, Liangliang Ji$^{6,7}$, Yitong Wu$^{6,8}$, Xuesong Geng$^{6,8}$, and  Markus B\"uscher$^{2,3}$}

	\affiliation{$^{1}$Institut f\"ur Theoretische Physik I, Heinrich-Heine-Universit\"at D\"usseldorf, Universit\"atsstr. 1, 40225 D\"usseldorf, Germany}
	
	\affiliation{$^{2}$ Peter Gr\"unberg Institut (PGI-6), Forschungszentrum J\"ulich, Wilhelm-Johnen-Str. 1, 52425 J\"ulich, Germany}
	
	\affiliation{$^{3}$ Institut f\"ur Laser- und Plasmaphysik, Heinrich-Heine-Universit\"at D\"usseldorf, Universit\"atsstr. 1, 40225 D\"usseldorf, Germany}
	
	\affiliation{$^{4}$ JARA-FAME (Forces and Matter Experiments), Forschungszentrum J\"ulich and RWTH Aachen University, 52056 Aachen, Germany}
	
	\affiliation{$^{5}$ Institut f\"ur Kernphysik, Forschungszentrum J\"ulich, Wilhelm-Johnen-Str. 1, 52425 J\"ulich, Germany}
	
	\affiliation{$^{6}$ State Key Laboratory of High Field Laser Physics, Shanghai Institute of Optics and Fine Mechanics, Chinese Academy of Sciences, Shanghai 201800, China}
	
	\affiliation{$^{7}$ CAS Center for Excellence in Ultra-intense Laser Science, Shanghai 201800, China}
	
	\affiliation{$^{8}$ Center of Materials Science and Optoelectronics Engineering, University of Chinese Academy of Sciences, Beijing 100049, China}

	\begin{abstract}
		The acceleration of polarized electrons and protons in strong laser and plasma fields is a very attractive option to obtain polarized beams in the GeV range. We investigate the feasibility of particle acceleration in strong fields without destroying an initial polarization, taking into account all relevant mechanisms that could cause polarization losses, i.e.\,the spin precession described by the T-BMT equation, the Sokolov-Ternov effect and the Stern-Gerlach force. Scaling laws for the \mbox{(de-)}polarization time caused by these effects reveal that the dominant polarization limiting effect is the rotation of the single particle spins around the local electromagnetic fields. We compare our findings to test-particle simulations for high energetic electrons moving in a homogeneous electric field. For high particle energies  the observed depolarization times are in good agreement with the analytically estimated ones.
	\end{abstract}
	
	\maketitle
	
	\section{Introduction}
	Spin-polarized particle beams are widely used for scattering experiments in nuclear and particle physics to study the structure and interaction of matter, to test the standard model \cite{EDM_2013, G-2_2015, Androic_2018}, to explore the structure of the proton and of QCD as our standard theory of the strong interaction \cite{Burkardt_2010},  to probe the nuclear structure \cite{COMPASS_2005}, or to investigate the dynamics of molecules \cite{Gay2009, Bederson_2017}. Up to now, conventional particle accelerators are used in most cases to prepare these beams and to deliver them to the experiments \cite{COSY, RHIC}.
	
	The technique of producing polarized beams strongly depends on the type of particle and their energies. For stable ones, such as electrons or protons, polarized sources can be employed with subsequent acceleration in a linear accelerator or a synchrotron. For instable particles like muons, polarization dependent particle decays are utilized \cite{G-2_2015}, while stable secondary beams, like antiprotons, might be polarized in dedicated storage rings by spin-dependent interactions \cite{Antiprotons_2005}. Electron or positron beams can be spontaneously polarized in magnetic fields of storage rings due to the emission of spin-flip synchrotron radiation acting on individual beam particles, the so-called Sokolov-Ternov effect in storage rings \cite{Sokolov1971, Mane_2005}. During acceleration of polarized beams in conventional circular accelerators, depolarizing spin resonances must be compensated by the use of complex correction techniques maintaining the beam polarization \cite{Mane_2005}. In linear accelerators, such a reduction of polarization can be neglected due to the very short interaction time between particle bunches and accelerating fields. 
	
	All methods mentioned above still rely on conventional particle accelerators that are typically very large in scale and budget. Therefore, concepts based on laser-driven wakefield acceleration or via ponderomotive scattering from extremely intense laser pulses have strongly been promoted during the last decades. The ultimate goal is to build the next generation highly compact and cost-effective accelerator facilities using a plasma as the accelerating medium, see e.g.\,Ref.\,\cite{EUPRAXIA_2017}. However, despite many advances in the understanding of fundamental physical phenomena, one largely unexplored issue here is how the particle spins are influenced by the huge electromagnetic fields which are inherently present in the plasma and what fundamental mechanisms may lead to the production of highly polarized beams. In general, there are two possible scenarios: either the magnetic field can align the spin of the accelerated beam particles, or the spins are too inert, so that a short acceleration has no influence on the spin alignment of a pre-polarized target. In this case, the polarization would be maintained throughout the whole acceleration process but a pre-polarized target would be required.
	\begin{figure}[t]
		\centering
		\includegraphics[width=\linewidth]{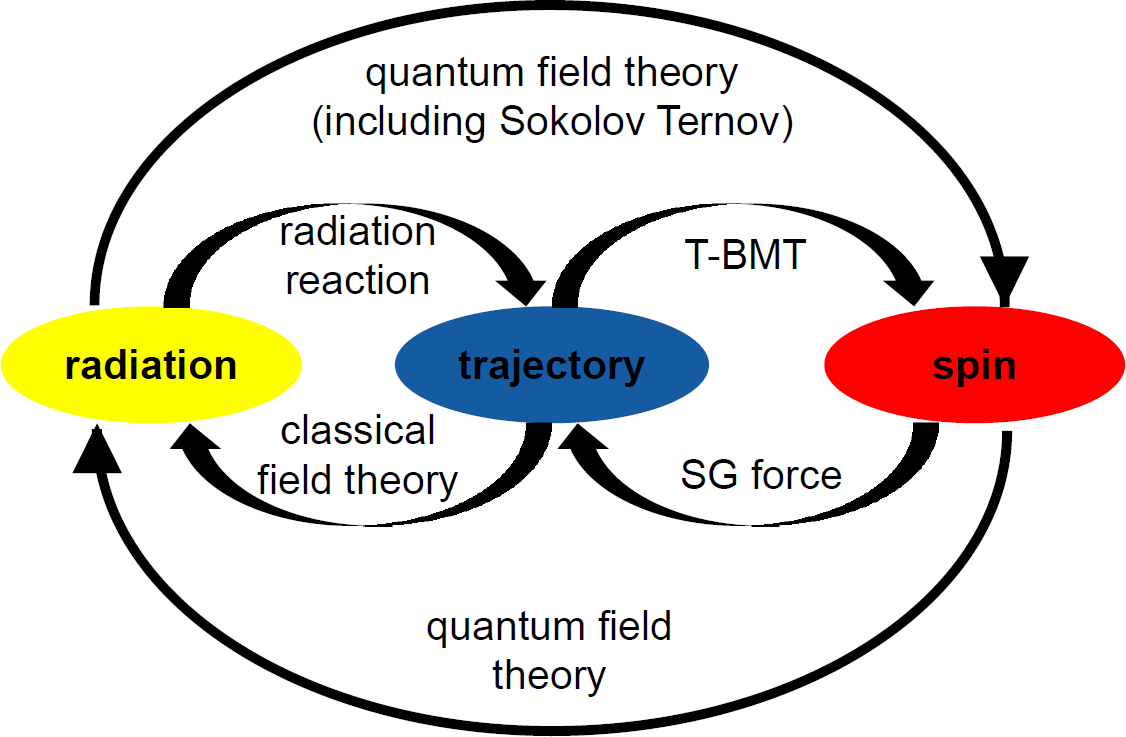}	
		\caption[overview]{Schematic overview of relations between single particle trajectory (blue), spin (red) and radiation (yellow). The black arrows indicate the basic physical process that links two of these fields.}
		\label{fig:overview}
	\end{figure} 
	
\begin{figure}[t]
		\centering
		\includegraphics[width=\linewidth]{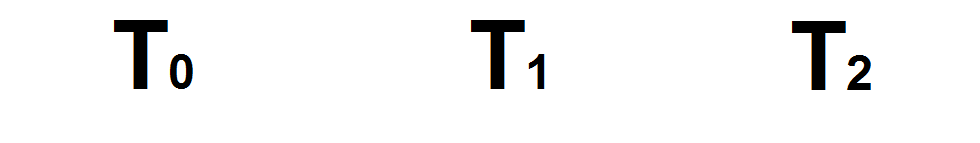}	\\
		\includegraphics[width=\linewidth]{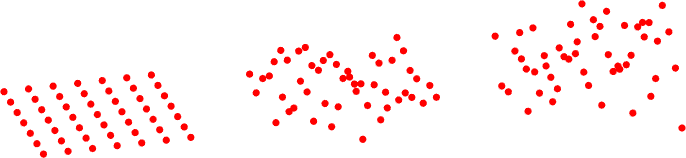}\\
		\includegraphics[width=\linewidth]{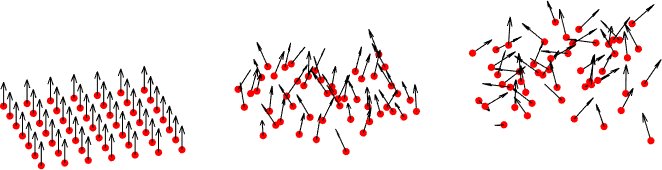}\\
		\includegraphics[width=\linewidth]{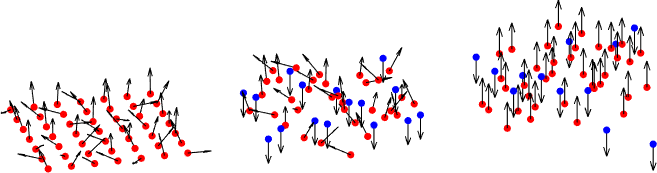}\\
		\includegraphics[width=\linewidth]{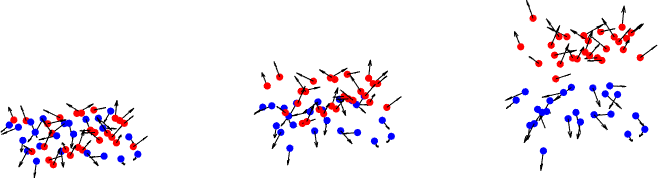}			
		\caption{Sketch of basic processes that need to be discussed in a laser-plasma accelerator for polarized particle beams. \textbf{1st line:} Pure kinetic evolution of an initially ordered particle ensemble in strong fields. \textbf{2nd line:} Same kinetic evolution as in the first line for an initially polarized particle beam. The black arrows indicate the single particle spins which are treated in a semi-classical limit following the T-BMT equation. \textbf{3rd line:} The particles' positions evolve as in the previous lines but their initially unordered spins individually flip in z-direction (red dots for spin up and blue dots for spin down) such that a certain polarization builds up. The corresponding mechanism is known as the Sokolov-Ternov effect. \textbf{Last line:} If the Stern-Gerlach force (or its mean) acts constantly for a longer time on an initially unpolarized and spatially unordered ensemble, we expect a certain beam split-up such that two polarized beams (red dots for spin up and blue dots for spin down) emerge.}
		\label{fig:teilchen}
	\end{figure}

	To characterize the actual scenario, a constant description of the relevant physical mechanisms is necessary. For example, if the single particle spins are treated in a semi-classical limit, it is the T-BMT equation (see section \ref{T-BMT}) that  determines the spin precession around the local electromagnetic field lines in dependence of the single particles' motion. If other physical effects are considered, it is important to know how they are related to the particle motion and the temporal spin evolution. Therefore, a schematic overview of the relevant relations between single particle trajectory (blue), spin (red) and radiation (yellow) is given in Fig.\,\ref{fig:overview}. Here, we see that the Stern-Gerlach force primary effects the trajectory of a particle, but only a theory also including the T-BMT equation self-consistently describes the particle and spin motion in electromagnetic fields. If the particles' energy is high enough, radiation effects must be considered, too. In the classical and semi-classical limit, acceleration of charged particles is treated within the framework of the classical field theory. This theory also describes the back action of radiation on the particle motion due to the radiation reaction force. In general, the radiation reaction force exceeds the Stern-Gerlach force by far if the particles are ultra-relativistic. But there are some field configurations which reverse this situation, so that the radiation reaction force can be neglected against the Stern-Gerlach force (see e.g.\,\cite{Flood2015}). A direct coupling between single particle spins and radiation fields is treated in the context of quantum field theory. In this theory, the mechanism describing the spontaneous self-polarization of an accelerated particle ensemble is known as the Sokolov-Ternov effect (also see Fig.\,\ref{fig:teilchen}, line\,3). Another possible polarization effect, that is not related to this rather quantum mechanical mechanism, is the possibility to spatially separate two overlapping, opposite polarized beams due to a constantly acting Stern-Gerlach force (also see Fig.\,\ref{fig:teilchen}, line\,4). A system, which includes the T-BMT spin motion solely, will only be able to describe the depolarization of an initially polarized ensemble based on an asynchronous spin precession (also see Fig.\,\ref{fig:teilchen}, line\,2).
	
	In our present work we follow Ehrenfest's theorem and treat the particle spin in a semi-classical approach such that the vector $\sbb$ can be interpreted as the expectation value $\sbb=\langle\Psi|\sigma|\Psi\rangle$ (here $|\Psi\rangle$ is the normalized two-component spinor and $\sigma$ are the Pauli matrices) \cite{Mane_2005}. Within this approach the polarization
	\begin{align}
		\Pb = \frac{1}{N}\sum_{i=1}^{N} \sbb_\mathrm{i} \label{POL}
	\end{align}	
	of an $N$-particle system is conserved relative to a certain axis, if the action angle 
	\begin{align}
		\alpha_\mathrm{max}=\max_i\frac{\Pb_\mathrm{0}\cdot\sbb_{\mathrm{i,f}}}{|\Pb_\mathrm{0}|\cdot|\sbb_{\mathrm{i,f}}|} \label{angle}
	\end{align}
	between the initial polarization $\Pb_\mathrm{0}$ and the finial spin vectors $\sbb_{\mathrm{i,f}}$ stays small. The processes relevant for polarization effects in conventional accelerators have been scrutinized in numerous works, see e.g.\,the Review by Mane et al. \cite{Mane_2005}. In analogy to these discussions the aim of our work is to identify the relevant mechanisms (see also Fig.\,\ref{fig:teilchen}) that may have an influence on beam polarization in laser-plasma accelerators, i.e.\,the T-BMT equation, the Sokolov-Ternov effect and the Stern-Gerlach force. 
	We derive rather general approximations of the strength of these effects that do not depend on the specific field configuration. Hence, it is the field gradient and the field strength from which we determine whether polarization conserving particle acceleration is possible or not. This approach allows us to analytically formulate scaling laws for the (de-)polarization time in fields which are as strong as those in a laser or wakefield accelerator for light electrons and comparatively heavy protons. Test-particle simulations for relativistic electrons moving in a homogeneous electric field  show that the analytically estimated minimum (de-)polarization time is in the same order as the observed one. Our numerical approaches include radiation (back-)reaction on the single particle trajectories. However, our analytical estimations are only valid as long as other quantum mechanical effects such as pair creation and emission of hard photons can be neglected.
	
	To model the spin precession, we solve the T-BMT equation analytically for various limiting cases and discuss the depolarization time as a function of the particle rest mass $m$, its energy $E = \gamma mc^2$ and the external field strength in section \ref{T-BMT}. In section \ref{SG} we estimate the perturbation of single particle trajectories in the context of a Lagrangian theory for the relativistic generalization of the Stern-Gerlach force. After a further generalization of the SG Lagrangian we discuss the Sokolov-Ternov effect in the scope of a Hamiltonian theory in section \ref{ST}. In the last section \ref{Test} we compare our findings to test-particle simulations for high energetic electrons moving in a homogeneous electric field. These simulations solve the equations of motion including radiation reaction and the Stern-Gerlach force. The observed depolarization times are in good agreement with the analytically estimated ones from section \ref{T-BMT}.

	\section{T-BMT}\label{T-BMT}
	In this section we derive the depolarization time $T_\mathrm{D}$ for an initially fully polarized electron or proton ensemble. The situation we describe is visualized in the second line of Fig.\,\ref{fig:teilchen}, where the particles (red dots) move independently of their individual spin vectors (black arrows) so that an initially (spatial and spin) ordered system state becomes unordered after a certain time (denoted by bold T). We focus on a situation where the single particle spins do not synchronize, such that the estimated depolarization time is a lower limit for laser-plasma accelerators. 
	
	If particles with mass $m$, charge $q$, anomalous magnetic moment $a$ and velocity $\vb$ move in an electromagnetic field $\Eb$, $\Bb$  with vanishing gradient, their spin vectors $\sbb_\mathrm{i}$ precess around the local electromagnetic fields and evolve according to the T-BMT equation
	\begin{align}
		\frac{d\sbb_\mathrm{i}}{dt} = -\mathbf{\Omega}\times\sbb_\mathrm{i}. \label{PugaSpinNorm}
	\end{align}	
	In cgs units the rotation frequency is simply \cite{Mane_2005}
	\begin{align}
		\mathbf{\Omega} = \frac{q}{mc}\left[\Omega_\mathrm{B}\Bb -\Omega_\mathrm{v}\left(\frac{\vb}{c}\cdot\Bb\right)\frac{\vb}{c} -\Omega_\mathrm{E}\frac{\vb}{c}\times\Eb\right],
	\end{align}	
	where 
	\begin{align}
		\Omega_\mathrm{B} = a+\frac{1}{\gamma}, && \Omega_\mathrm{v} = \frac{a \gamma}{\gamma+1}, && \Omega_\mathrm{E} = a +\frac{1}{1+\gamma}.
	\end{align}	
	If all spin vectors in the N-particle ensemble precess coherently, a certain polarization changes its orientation in space but its absolute value is conserved. This would be possible if all particles saw the same electromagnetic field and if all particles moved on similar trajectories. Particularly, the conservation of polarization during laser-plasma acceleration has to be considered in two parts: (i) during the injection of low energetic ($\gamma\approx1$) particles into the (laser- or wake-) field; (ii) during the steady acceleration phase of already relativistic ($\gamma\gg1$) particles. In the following, we focus on the theoretical description of phase (ii) by deriving the depolarization time from the T-BMT equation solely. For a detailed analysis of (i) we refer to the work of Wu et. al \cite{Wu2019} and Wen et. al \cite{Wen2019}. Here, the conservation of polarization during the injection of electrons into a wakefield, driven by an intense laser pulse, is demonstrated and explained. 
	
	In this work, we derive a scaling law for the minimum depolarization time for electrons (and positrons) and protons in strong fields without discernible symmetry. Precise and case dependent estimations for the depolarization time can be made for an exactly known local field distribution. For example, if we considered the highly symmetric potentials known from spherical, quasi-static bubble and blow-out models \cite{Kostyukov2009, Kostyukov2010, Lu2006, Lu2006a, Kalmykov2009, Kalmykov2011, Yi2011, Yi2013, Pak2010, Zeng2012, Thomas2014}, we could solve the T-BMT equation analytical and would find that the polarization of an initially zero emittance electron bunch is preserved for the whole acceleration time \cite{Vieira2011}. If we substituted the accelerating and focusing fields from more general models for channeled plasma \cite{Thomas2016, Golovanov2017b, Golovanov2016, Golovanov2016b}, we could study the influence of the plasma density profile on the polarization evolution analytically.
	
	If the field configuration is rather unsymmetrical, it is hard to decide whether a given polarization is conserved. In this case, the easiest way to estimate a minimum for the depolarization time $T_\mathrm{D}$, is to neglect any symmetry effects and to take only the total field strength, which is as high as in plasma accelerators, into account. Before we start, we simplify the T-BMT equation by normalizing the system variables to elementary charge $e$, electron mass $m_\mathrm{e}$, speed of light $c$, momentum $m_\mathrm{e}c$, energy $m_\mathrm{e}c^2$, spin $\hbar/2$, time $\omega_\mathrm{L}^{-1}$, lengths $k_\mathrm{L}^{-1}$, (laser) field strengths $E_\mathrm{0}=m_\mathrm{e}c\omega_\mathrm{L}/e$, and (critical) density $n_\mathrm{c}=m_\mathrm{e}\omega_\mathrm{L}^2/(4\pi e^2)$. With this normalization we refer to a laser with frequency $\omega_\mathrm{L}$ and wavelength $\lambda_\mathrm{L}$ which interacts with the polarized particle ensemble. Then the normalized T-BMT frequency for a particle with mass $m$ and anomalous magnetic moment $a$ becomes
	\begin{align}
		\mathbf{\Omega} = q\frac{m_\mathrm{e}}{m}\left[\Omega_\mathrm{B}\Bb -\Omega_\mathrm{v}\left(\vb\cdot\Bb\right)\vb -\Omega_\mathrm{E}\vb\times\Eb\right]. \label{omegaNorm}
	\end{align}
	To estimate whether a given polarization $\Pb$ is preserved during the time $T$, we need to solve Eq.(\ref{PugaSpinNorm}) for each particle. However, for an analytical estimation it is sufficient to know the maximum action angle $\alpha_\mathrm{max} $ (see Eq.\,(\ref{angle})) between the initial polarization $\Pb_\mathrm{0}$ and the finial spin vectors $\sbb_{\mathrm{i,f}} = \sbb_\mathrm{i}(T)$. If the individual spin precessions are rather incoherent, the initial polarization direction is conserved but its absolute value would be decreased to $P(T) \geq P_\mathrm{0} -\sin(\alpha_\mathrm{max})$. As soon as $\alpha_\mathrm{max}$ is in the order of $\pi/2$, we expect that the N-particle ensemble is completely depolarized and call the corresponding time the minimum depolarization time $T_\mathrm{D}$.
	
	The anomalous magnetic moment of an electron is $a_\mathrm{e}=\alpha/(2\pi)\approx10^{-3}$, where $\alpha$ is the fine structure constant. For a proton $a_\mathrm{p}\approx a_\mathrm{e}m_\mathrm{p}/m_\mathrm{e}$, so that $\mathrm{d}\sbb/\mathrm{d}t$ is equal for protons and electrons in the relativistic limit $\gamma\gg10^3$ (a more precise justification is given in the following). In general, ions with mass $m_\mathrm{I}\approx m_\mathrm{p}\gg m_e$ have an anomalous magnetic momentum $a_\mathrm{I}\gg a_\mathrm{e}$ so that we can expect that they behave like protons if their gamma factor is much lager than one. To determine, how much a given polarization changes, or for a more precise estimation of $T_\mathrm{D}$, we assume a relativistic $(|\vb|\approx c)$ ensemble in which all particles have nearly the same energy but move in different directions, so that the single spin precession axes are not aligned. Since the rotation around the $\mathbf{\omega}$-axis can be interpreted as a superposition of precessions around the $\mathbf{B}$-, $\mathbf{v}$- and $\mathbf{v}\times\mathbf{E}$-axis, a substantial approximation of the absolute value of the precession frequency for this situation is
	\begin{align}
		|\mathbf{\Omega}| \leq \frac{m_\mathrm{e}}{m}( \Omega_\mathrm{B}|\Bb| + \Omega_\mathrm{v}|\Bb| + \Omega_\mathrm{E}|\Eb| ), \label{AO1}
	\end{align}
	where equality holds if all precession axes were aligned. To calculate a lower limit for $T_\mathrm{D}$, we estimate an upper limit of $|\mathbf{\Omega}|$ by substituting the dominant field strength $F=\max(\Eb,\Bb)$ for $|\Eb|$ and $|\Bb|$. Then 
	\begin{align}
		|\mathbf{\Omega}| < \frac{m_\mathrm{e}}{m}\left[\Omega_\mathrm{B} + \Omega_\mathrm{v} + \Omega_\mathrm{E}\right]F, \label{AO2}
	\end{align}	
	which depends on the maximum field strength $F$, the particle mass $m$ and the ratio of $a$ and $\gamma$ solely. For relativistic electrons the Omega terms can be summarized to $\Omega_\mathrm{B} + \Omega_\mathrm{v} + \Omega_\mathrm{E}\approx 3a_\mathrm{e}+2/\gamma$, while for electrons with energy of more than 10 GeV it is $\gamma\gg1/a_\mathrm{e}$ so that
	\begin{align}
		\Omega_e \approx 3a_\mathrm{e}F.
	\end{align}	
	This implies that the spin evolution in unsymmetrical fields is independent of the electron energy for sufficiently high energies. 
	
	For relativistic $\gamma\gg1$ protons with an energy of a few hundred GeV and above the $2/\gamma$ term can be neglected because of the large anomalous magnetic moment $a_\mathrm{p}>1$ and we immediately find
	\begin{align}
		\Omega_{\mathrm{p,TeV}} = \Omega_{\mathrm{e,GeV}}.
	\end{align}	 
	In short words one can say that the (near) TeV-proton spin motion is equivalent to that of a GeV-electron if the fields do not exhibit a certain symmetry. If the fields were symmetric, e.g. like those in a circular accelerator, one would find another relation which would take other effects like the orbital motion in rings into account \cite{Courant1980}.
	
	For lower proton energies in the GeV-regime such that $\gamma$ is in the range of unity but still $|\vb|\approx 1$
	\begin{align}
		\Omega_{\mathrm{p,rel}} \approx \Omega_{\mathrm{e,GeV}} + \frac{m_\mathrm{e}}{m_\mathrm{p}}\left(\frac{1}{\gamma} + \frac{1-a_p}{1+\gamma}\right)F. \label{AO3}
	\end{align}	
	For protons $a_\mathrm{p}\approx 1.8$ such that the second term slightly increases $\Omega_{\mathrm{p,rel}}$ to
	\begin{align}
		\Omega_{\mathrm{p,GeV}} \approx a_\mathrm{e}F\left(3 + \frac{0.6}{a_\mathrm{p}}\right) \approx 3.3a_\mathrm{e}F.\label{AO4}
	\end{align}	
	This limit is only ten percent larger than the previously found one and converges quite fast. For other ions with smaller anomalous magnetic moment than $a_\mathrm{p}$, the spin frequency must be calculated from Eq.(\ref{AO2}) and can be much smaller than $3.3a_\mathrm{e}F$. Especially low energetic ions with negative anomalous magnetic moment the factor $3a+2/\gamma$ can be almost zero. For this reason we interpret Eq.(\ref{AO4}) as the only relevant limit for our work and find the electron and proton minimum depolarization time
	\begin{align}
		T_{\mathrm{D,p}}  = \frac{\pi}{6.6a_\mathrm{e}F}, \label{AO5}
	\end{align}	
	which is independent of the particles' energy.
	
	Since we are interested in particle acceleration in strong fields which are in the same order as those in laser-plasma accelerators, we have to interpret $E_\mathrm{0}=m_\mathrm{e}c\omega_\mathrm{L}/e$ (which we used for variable normalization) as a laser field. Then we can expect that $F$ is in the order of unity if the particle beam should be separated by the laser itself. The plasma fields are usually $\epsilon=\omega_\mathrm{p}/\omega_\mathrm{L}$-times smaller than $E_\mathrm{0}$ \cite{Mora1997}. Thus, for polarization conservation in fields which are as strong as those in a plasma wakefield, we have to substitute $F\equiv\epsilon$. Substituting $F=1$ leads to $T_{\mathrm{D,L}}\approx 520 \omega_\mathrm{L}^{-1}$ both for relativistic ions and ultra-relativistic electrons. The lower limit of the depolarization time for electrons and protons in a wakefield is in the range of $T_{\mathrm{D,W}}\approx 520 \omega_\mathrm{L}^{-1}\epsilon^{-1}\gg T_{\mathrm{D,L}}$. In the first case, $T_{\mathrm{D,L}}$ is in the range of pico-seconds - a time span large enough to guaranty polarization conservation during the interaction with passing ps laser pulses. In the scope of wakefield acceleration in underdense plasma, the second time corresponds to an acceleration length in the mm range.
	
	In section \ref{Test} we compare the analytically derived depolarization time in Eq.(\ref{AO5}) to the numerically found one from test-particle simulations of a high energetic non-zero-emittance electron beam moving in a homogeneous electric field. By varying the initial momentum spread we observe that the depolarization times scale as predicted by Eq.\,(\ref{AO5}). However, Eq.\,(\ref{AO5}) must still be interpreted as a lower limit.
	
	Two important examples of conserved polarization during acceleration are TNSA with pre-polarized targets \cite{Huetzen2019, Huetzen2019_a, Buescher2019} and wakefield acceleration in pre-polarized gases \cite{Wu2019, Wu2019_a}. In the latter case both electrons and protons could be accelerated without significant change in polarization. In the following chapters we show why it is sufficient to solve the T-BMT equation, without considering further quantum effects like the Stern-Gerlach force or self-polarization due to spin-flips.
	
	\section{Stern-Gerlach}
	\label{SG}
	In this chapter, we estimate the spin back-action on the particle trajectory by approximating the relativistic generalization of the Stern-Gerlach [SG] force for fast ($|\vb|\approx c$) but uncorrelated moving charged particles with spin 1/2. The situation we describe is visualized in the last line of Fig.\,\ref{fig:teilchen}, where the particles (blue and red dots) move according to their individual spin vectors (black arrows) either in positive or negative (z-)direction, if the SG force acted constantly in $z$-direction. In this way an initially (spatial and spin) disordered system state separates into two independent particle sets with only positive (red dots with arrow up) or negative (blue dots with arrow down) $s_\mathrm{z}$ component allowing for a certain polarization to be built up. For the case of highly symmetric, quasi-static fields we solve the equations of motion for test particles numerically in section\,\ref{Test}. 
	
	To find an expression for the SG force in dependence of the spin precession we start with the Lagrangian 
	\begin{align}
		L_\mathrm{total}=L_\mathrm{EM}+L_\mathrm{SG}. \label{Ltot1}
	\end{align}  
	Here
	\begin{align}
		L_\mathrm{EM} = -\frac{mc^2}{\gamma} + \frac{q}{c}\vb\cdot\Ab - q\varphi
	\end{align}
	is the Lagrangian describing the motion of the particle in the external electromagnetic field $\Eb = -\nabla\varphi -1/c \cdot \partial\Ab/\partial t$, $\Bb=\nabla\times\Ab$ and 
	\begin{align}
		L_\mathrm{SG} = -\mathbf{\Omega}\cdot\sbb
	\end{align}
	is the spin interaction Lagrangian \cite{Pomeranskii1998}. With this ansatz the canonical momentum becomes  
	\begin{align}
		\Pb = \gamma m\vb +\frac{q}{c}\Ab + \Pb_\mathrm{SG},
	\end{align}
	where the Stern-Gerlach part is simply 
	\begin{align}
		\Pb_\mathrm{SG} = \nabla_\mathrm{\vb} L_\mathrm{SG}.
	\end{align}
	From the Lagrange equation of motion we finally find
	\begin{align}
		\frac{d(\gamma m\vb)}{dt} = \frac{d\pb_\mathrm{kin}}{dt} = \Fb_\mathrm{total} = \Fb_\mathrm{EM} + \Fb_\mathrm{SG}
	\end{align}
	with the Lorentz force
	\begin{align}
		\Fb_\mathrm{EM} = q\Eb + \frac{q}{c}\vb\times\Bb \label{LF}
	\end{align}
	and the Stern-Gerlach force
	\begin{align}
		\Fb_\mathrm{SG} = \left(\nabla  -\frac{d}{dt}\nabla_\mathrm{\vb}\right) (\mathbf{\Omega}\cdot\sbb). \label{FSG3}
	\end{align}
	If we apply the same normalization as in the beginning of section \ref{T-BMT} including a spin normalization to $\hbar/2$, this expression changes to 
	\begin{align}
		\Fb_\mathrm{SG} = \Lambda_\mathrm{SG}\left(\nabla  -\frac{d}{dt}\nabla_\mathrm{\vb}\right) (\mathbf{\Omega}\cdot\sbb), \label{FSG2}
	\end{align}
	where $\mathbf{\Omega}$ has to be taken from Eq.(\ref{omegaNorm}), $s=1$ and 
	\begin{align}
		\Lambda_\mathrm{SG} = \frac{\hbar\omega_\mathrm{L}}{2m_\mathrm{e}c^2} \approx 1.2\cdot10^{-6}\lambda_\mathrm{L}[\mu m]^{-1}
	\end{align}
	is the ratio between the energy of a photon with wavelength $\lambda_\mathrm{L}$ and the electron rest energy. If we considered the SG force in fields which are as strong as wakefields $\Lambda_\mathrm{SG}$ would be in the order of $10^{-7}$ to $10^{-8}$. If the fields were as strong as those in a laser with a few tens of nm wavelength, $\Lambda_\text{SG}$ would be in the order of $10^{-4}$ to $10^{-5}$.
	
	For an analytical estimation of how far the SG force might change the total polarization of a relativistic particle (electron, proton, ect.) ensemble, we proceed in the same way we did in the previous section, i.e. we neglect all field symmetries and take only the dominant field strength $F=\max(\Eb,\Bb)$ and the dominant field gradient $\partial F = \max_{\mathrm{i,j}}(|\partial_{\mathrm{x_i}}F_\mathrm{j}|)$ into account (for more information see appendix). Then we estimate an upper limit for $|\Fb_\text{SG}|$ and discuss in how far this force could build up a certain polarization in an initially unpolarized system due to a separation of particles with opposite spins.
	
	In the non-relativistic ($|\vb|\ll 1$) limit, the first term of $\Fb_\mathrm{SG}$ is the well known Stern-Gerlach force
	\begin{align}
		\Fb_\mathrm{SG} = \Lambda_\mathrm{SG}\frac{qm_\mathrm{e}}{m}(1+a)\nabla(\Bb\cdot\sbb) = \nabla(\mathbf{\mu}\cdot\Bb). \label{kth}
	\end{align}
	For relativistic particles, the correction in $\Fb_\mathrm{SG}$ does not consist of the $\mathrm{d}/\mathrm{d}t\nabla_\mathrm{\vb}$-term solely but also includes contributions from $\Omega_\mathrm{v}\nabla$ and $\Omega_\mathrm{E}\nabla$. They can easily be written down if we use the vector identities from Appendix \ref{APP_SG}  and if we treat the coordinates $\rb$, the canonical momentum $\Pb$ and the spin $\sbb$ as independent variables. Then the coefficients $\Omega_\mathrm{B}$, $\Omega_\mathrm{v}$ and $\Omega_\mathrm{E}$ depend on $|\vb|$ and the gradient acts on the fields solely. The relevant terms we need for the gradient are $\nabla(\Bb\cdot\sbb)$, $\nabla(\Bb\cdot\vb)$ and $\nabla(\Eb\times\vb)$. They cannot be further simplified but, since the normalized spin has length one, their norm is limited by $3\sqrt{3}\partial F$ (see \ref{grad_Fa}). Thus, with the same arguments as in section \ref{T-BMT}, it is clear that for $\gamma\gg1$ the contribution of spatial gradient to $\Fb_\text{SG}$ is limited by
	\begin{align}
		F_\nabla = \Lambda_\mathrm{SG}\frac{qm_\mathrm{e}}{m}\left(3a+\frac{2}{\gamma}\right)3\sqrt{3}\partial F. \label{Fnab}
	\end{align}
	
	The second term in Eq.(\ref{FSG2}) contributes due to temporal (and spatial) field variations (T part), the energy changing rate (E part) and the change in direction (V part) of the accelerated particle. With Eq.(\ref{PugaSpinNorm}) its k-th component can be written as
	\begin{align}
		\frac{d}{dt}\frac{\partial}{\partial v_\mathrm{k}}(\mathbf{\Omega}\cdot\sbb) = \frac{d}{dt}\frac{\partial\mathbf{\Omega}}{\partial v_\mathrm{k}}\cdot\sbb  -\frac{\partial\mathbf{\Omega}}{\partial v_\mathrm{k}}\cdot(\mathbf{\Omega}\times\sbb), \label{dk}
	\end{align}	
	where
	\begin{align}
		\frac{m}{qm_\mathrm{e}}\frac{d}{dt}\frac{\partial \mathbf{\Omega}}{\partial v_\mathrm{k}} = T_\mathrm{k} + E_\mathrm{k} + V_\mathrm{k}.	\label{dt1}
	\end{align}	
	Before we calculate these three parts in detail, it is important to mention that we are interested in an upper limit for the norm of the second derivative in Eq.(\ref{FSG2}) solely. In turn, this limit predominantly depends on the leading term, so that it is sufficient to identify the dominant terms and consider noting but their proportionality. Since all derivations are symmetric in $k$, it is convenient to calculate the T, E, and V terms for one (arbitrary) $k$ only.
	
	Since \Eb and \Bb depend on \rb and $t$ solely and since $\Omega_\mathrm{B}$, $\Omega_\mathrm{v}$ and $\Omega_\mathrm{E}$ are functions with sole argument $\gamma$, the first order derivation in Eq.(\ref{kth}) reduces to
	\begin{align}
		\frac{m}{qm_\mathrm{e}}\frac{\partial \mathbf{\Omega}}{\partial v_\mathrm{k}} =& \left[\Omega_\mathrm{B}'\Bb -\Omega_\mathrm{v}'\left(\vb\cdot\Bb \right)\vb -\Omega_\mathrm{E}' \vb\times\Eb \right] \gamma^3 v_\mathrm{k} \nonumber\\
		& -\Omega_\mathrm{v} B_\mathrm{k}\vb -\Omega_\mathrm{v}\left(\vb\cdot\Bb\right)\hat{e}_\mathrm{k} -\Omega_\mathrm{E}\hat{e}_\mathrm{k}\times\Eb \label{dvk}
	\end{align}	
	where we have used the abbreviation $\Omega_\mathrm{\alpha}'=\mathrm{d}\Omega_\mathrm{\alpha}/\mathrm{d}\gamma$ for $\alpha=B,v,E$ and $\partial\gamma/\partial v_\mathrm{k} = \gamma^3 v_\mathrm{k}$.
	Since we are interested in relativistically moving particles, we assume $|\vb|\approx1$ and $\gamma\gg1$. Then - after a separation of the total time derivation into the spatial part $\partial_\mathrm{X} = \partial/\partial t +\vb\cdot\nabla $ and the velocity-energy part $\dot{\vb}\cdot\nabla_\mathrm{\vb} = \dot{\gamma}\partial_\mathrm{\gamma} = q(\vb\cdot\Eb)\partial_\mathrm{\gamma}$ (see Appendix \ref{dt} to \ref{vdot}) - the second order derivation in Eq.(\ref{dk}) can be split up into the temporal (and spatial) part
	\begin{align}
		T_\mathrm{K}= & \left[\Omega_\mathrm{B}'\partial_\mathrm{X}\Bb -\Omega_\mathrm{v}'\left(\vb\cdot\partial_\mathrm{X}\Bb \right)\vb -\Omega_\mathrm{E}' \vb\times\partial_\mathrm{X}\Eb \right] \gamma^3 v_\mathrm{k} \nonumber\\
		- & [\Omega_\mathrm{v} \partial_\mathrm{X} B_\mathrm{k}\vb +\Omega_\mathrm{v}\left(\vb\cdot\partial_\mathrm{X}\Bb\right)\hat{e}_\mathrm{k} +\Omega_\mathrm{E}\hat{e}_\mathrm{k}\times\partial_\mathrm{X}\Eb] ,\label{dtS}
	\end{align}	
	the energy part
	\begin{align}
		E_\mathrm{k} = & \left[\Omega_\mathrm{B}''\Bb -\Omega_\mathrm{v}''\left(\vb\cdot\Bb \right)\vb -\Omega_\mathrm{E}'' \vb\times\Eb \right] \gamma^3 v_\mathrm{k} q(\vb\cdot\Eb) \nonumber\\
		+ & \left[\Omega_\mathrm{B}'\Bb -\Omega_\mathrm{v}'\left(\vb\cdot\Bb \right)\vb -\Omega_\mathrm{E}' \vb\times\Eb \right] 2\gamma^2 v_\mathrm{k} q(\vb\cdot\Eb) \nonumber\\
		- & [\Omega_\mathrm{v}' B_\mathrm{k}\vb +\Omega_\mathrm{v}'\left(\vb\cdot\Bb\right)\hat{e}_\mathrm{k} +\Omega_\mathrm{E}'\hat{e}_\mathrm{k}\times\Eb]q(\vb\cdot\Eb)
		,	\label{dtE}
	\end{align}	
	and the velocity (change in direction) part
	\begin{align}
		V_\mathrm{k} = & \left[ -\Omega_\mathrm{v}'\left(\dot{\vb}\cdot\Bb \right)\vb -\Omega_\mathrm{v}'\left(\vb\cdot\Bb \right)\dot{\vb} -\Omega_\mathrm{E}' \dot{\vb}\times\Eb \right] \gamma^3 v_\mathrm{k} \nonumber\\
		& + \left[\Omega_\mathrm{B}'\Bb -\Omega_\mathrm{v}'\left(\vb\cdot\Bb \right)\vb -\Omega_\mathrm{E}' \vb\times\Eb \right] \gamma^3 \dot{v}_\mathrm{k}   \nonumber\\
		& -\Omega_\mathrm{v} B_\mathrm{k}\dot{\vb} -\Omega_\mathrm{v}\left(\dot{\vb}\cdot\Bb\right)\hat{e}_\mathrm{k}.
		\label{dtV}
	\end{align}	
	
	To estimate, whether a bunch of charged particles with spin $\hbar/2$ can be separated by the SG-T force, it is necessary to find the dominant terms in Eq.(\ref{FSG2}). To do that, we first compare the norm of each term in $T_\mathrm{k}$, $E_\mathrm{k}$ and $V_\mathrm{k}$. For simplicity we substitute $F=\max(|\Eb|,|\Bb|)$, $|\partial_\mathrm{X}\Bb|=|\partial_\mathrm{X}\Eb|=4\partial F$ and $|\sbb|=1$. Then the spin frequency components scale like
	\begin{align}
		|\Omega_\mathrm{B}| & \approx |\Omega_\mathrm{E}| \approx a +\frac{1}{\gamma}, && |\Omega_\mathrm{v}| \approx a,  \\	
		|\Omega_\mathrm{B}'| &\approx |\Omega_\mathrm{E}'| \approx \frac{1}{\gamma^2}, && |\Omega_\mathrm{v}'| \approx \frac{a}{\gamma^2}, \\
		|\Omega_\mathrm{B}''| &\approx |\Omega_\mathrm{E}''| \approx \frac{2}{\gamma^3}, && |\Omega_\mathrm{v}''| \approx \frac{2a}{\gamma^3}
	\end{align}	
	such that the leading terms in Eqs.(\ref{dtS}) and (\ref{dtE}) are
	\begin{align}
		|T_\mathrm{k}| \leq 4(2+a)\gamma\partial F, && |E_\mathrm{k}| \leq 4(2+a)qF^2. \label{TE}
	\end{align}	
	To find the dominant terms in Eq.(\ref{dtV}) we also substitute $|\dot{\vb}|= 3qF/\gamma$ from Eq.(\ref{vdot}) and find
	\begin{align}
		|V_\mathrm{k}| & \leq 9(1 +a)qF^2. \label{V}
	\end{align}	
	If we apply Eq.(\ref{AO2}) and the same approximation as above to Eq.(\ref{dvk}) we further see that the spin-change-rate is limited by
	\begin{align}
		\left|\frac{\partial\mathbf{\Omega}}{\partial v_\mathrm{k}}\cdot(\mathbf{\Omega}\times\sbb)\right| \leq \frac{m_\mathrm{e}^2}{m^2}(2+a)(2 +3a\gamma) F^2. \label{partO}
	\end{align} 
	If we now compare all terms in Eq.(\ref{FSG2}) we see that there are four basic proportionalities we have to consider in the limit $\gamma\gg1$. The first one is related to the spatial gradient (see Eq.(\ref{Fnab})). The second proportionality is related to the spin-change-rate (see Eq.(\ref{partO})) and contributes with terms
	\begin{align}
		F_\mathrm{s} \propto \Lambda_\mathrm{SG}\frac{m_\mathrm{e}^2}{m^2}(2+3a\gamma)F^2. \label{Fs}
	\end{align}
	Very similar to this term is
	\begin{align}
		F_\mathrm{{EV}} \propto \Lambda_\mathrm{SG}\frac{m_\mathrm{e}}{m}F^2, \label{FEV}
	\end{align}
	a proportionality we see due to the energy changing rate and the change in direction. The last part of $\Fb_\mathrm{SG}$, we have to consider, is related to temporal and spatial field variations again. It scales like
	\begin{align}
		F_\mathrm{T} \propto \Lambda_\mathrm{SG}\frac{m_\mathrm{e}}{m}\gamma \partial F, \label{FS}
	\end{align}
	which is $\gamma$-times stronger than the gradient part $F_\nabla$. From this we immediately see that the SG-T part, which is related to the ordinary field gradient, is always overcompensated by the $\mathrm{d}/\mathrm{d}t\nabla_\mathrm{\vb}$-part - the field gradient part which must be considered if the particle moves relativistically.
	
	The prefactors, we did not take into account in the above proportionalities, are all of the same order. Thus, the ratio of the field strength, the field gradient and the $\gamma$ factor determines which of the terms Eqs.(\ref{Fs}) to (\ref{FS}) predominantly describes the back action of the spin on the trajectory. If we took all prefactors of $|T_\mathrm{k}|$, $|E_\mathrm{k}|$, $|V_\mathrm{k}|$, and Eq.(\ref{partO}) into account, we would see that all forces are up to one order of magnitude larger if $a>1$, i.e. in general, heavy ions feel an up to ten times stronger force than light electrons. In the following we identify the dominant term for electrons and protons and compare it to the electromagnetic force $F$. 
	
	For electrons it is $m_\mathrm{e}/m=1$ and $a=a_\mathrm{e}\ll1$, so that $F_\mathrm{s} \gg F_\mathrm{EV}$ for both GeV energies, where $1/a_\mathrm{e}^2\gg\gamma\gg1/a_\mathrm{e}$, and TeV energies, where $\gamma\gg1/a_\mathrm{e}^2\gg1$. As a consequence, an alternation of the trajectory due to the coupling of the spin to the acceleration can always be neglected against an alternation of the trajectory due to the spin-change-rate caused by the T-BMT rotation. This observation is perfectly in line with the experience gained at conventional particle accelerators \cite{Mane_2005}. If the fields are homogeneous and quasi-static we can assume $\partial F=0$, so that the T-BMT rotation is the only relevant mechanism. If, however, $\partial F \gg a_\mathrm{e}F^2$, it is the (temporal and spatial) field variation that separates two electron beams with opposite orientated spins. 
		
	For protons we have to consider the mass ration $m_\mathrm{e}/m_\mathrm{p} = a_\mathrm{e}/a_\mathrm{p} \ll1$ in the proportionalities Eqs.(\ref{Fs}) to (\ref{FS}), so that the above discussion for electrons must be reformulated. The first difference to the electronic case is visible for rather moderate $1/a_\mathrm{e}\gg\gamma\gg 1$ relativistic protons with energy in the lower GeV-regime. These protons see the force $F_\mathrm{EV}\gg F_\mathrm{s}$, which means that here the T-BMT rotation is overcompensated by the spin-acceleration coupling if $\partial F=0$. For $\partial F \gg F^2/\gamma$ it is the (temporal and/ or spatial) field variation that dominates all other forces. In the second (TeV or $\gamma\gg 1/a_\mathrm{e}$) case the situation changes back to that known for electrons so that $F_\mathrm{s} \gg F_\mathrm{EV}$ for $\partial F=0$. Similar to the electronic case, the (temporal and/ or spatial) field variation is the relevant mechanism if $\partial F \gg a_\mathrm{e} F^2$.
	
	Summarizing the electronic and the ionic case in the high-energy ($\gamma\gg1/a_\mathrm{e}$) regime we point out that the particle trajectories are perturbed rather due to spin T-BMT rotation than due to the energy- or velocity-changing rates. Further, both electrons and ions are sensitive to the field gradients, which means that, even for small field variations, the $F_T$ part of the SG-T force overcompensates all other effects. In general, for electrons the Stern-Gerlach force is much smaller than the electromagnetic force as long as $\Lambda_\mathrm{SG}a_\mathrm{e}\gamma F^2\ll F$ or $\Lambda_\mathrm{SG}\gamma\partial F\ll F$. For a field strength in the order of $F=1$ this means that a significant perturbation of single-particle trajectories must be considered as soon as $\gamma\approx(\Lambda_\mathrm{SG}\partial F)^{-1}$ in gradient fields, which is already achieved for electron energies in the range of a $100$\,GeV. For protons, a 2000-times larger kinetic energy would be necessary to see a trajectory perturbation due to the Stern-Gerlach force.
	
	Now that we have understood the principle mechanism in the SG force for high energetic particles, we want to estimate whether this force can be used to separate (or filter) an unpolarized relativistic particle beam such that, after having passed a specific acceleration length in strong electromagnetic fields, two polarized beams emerge. Since we are interested in particle acceleration in strong fields which are in the same order as those in laser-plasma accelerators, we have to interpret $E_\mathrm{0}$ as a laser field. Then $F\equiv1$ and $\partial F\equiv 1/(2\pi)$, if the particle beam should be separated by the laser itself. The plasma fields are usually $\epsilon=\omega_\mathrm{p}/\omega_\mathrm{L}$-times smaller than $E_\mathrm{0}$. Thus, for polarization in fields which are as strong as those in pure plasma, we have to substitute $F\equiv\epsilon$ and $\partial F\equiv\epsilon/R_\mathrm{L}$, where $R_\mathrm{L}\equiv2\pi/\epsilon$ is the laser focal spot radius (for more information see also Ref.\,\cite{Mora1997}). Then, if the SG-force acted constantly in one direction perpendicular to the beam propagation direction, we could assume that the particles' energy is conserved while the unpolarized beam would split up into two co-propagating polarized beams with spatial distance
	\begin{align}
		\Delta = \frac{m_\mathrm{e}}{m}|\Fb_\mathrm{SG}| T_\mathrm{acc}^2 \gamma^{-1}.
	\end{align}	
	Since we know the minimum depolarization times from the previous section, we assume $T_\mathrm{acc} \propto 500$ for the system specific acceleration time if the field strength is comparable to that from lasers ($F=1$) and $T_\mathrm{acc} \propto 500\epsilon^{-1}$ if it is comparable to that from wakefields ($F=\epsilon$). In this way the resulting distance $\Delta$ is independent from the specific field strength $F$ and from the smallness parameter $\epsilon$.
	
	If we consider particles with energy in the GeV-range in a tailored plasma channel \cite{Golovanov2016b, Thomas2016}, we can assume a certain field homogeneity ($\partial F = 0$), so that it is the $F_\mathrm{s}$-term that must be considered as a primary source for SG-force for electrons and the $F_\mathrm{EV}$-term that must be considered for protons. This in turn leads to a maximum particle separation distance of
	\begin{align}
		\Delta_\mathrm{e}(\partial F=0) \propto 0.3(2+3a\gamma) \lambda_\mathrm{L}[\mu m]^{-1}\gamma^{-1}
	\end{align}
	for electrons and 
	\begin{align}
		\Delta_\mathrm{p}(\partial F=0) \propto	0.3\left(\frac{m_\mathrm{e}}{m}\right)^2 \lambda_\mathrm{L}[\mu m]^{-1}\gamma^{-1}
	\end{align}
	for protons. These distances are in the nm-range for electrons and in the sub pm-range for protons. 
	
	For electrons, an additional field gradient must be considered if $\partial F\gg a_\mathrm{e}F^2\propto a_\mathrm{e}\epsilon^2$. Thus value is much smaller than the usual plasma field gradient $\partial F\equiv\epsilon^2/(2\pi)$ so that even small field and density perturbations almost immediately destroy the idealized $\partial F=0$ scenario for electrons. For protons, an additional field gradient must be considered if $\partial F\gg F^2/\gamma\propto \epsilon^2/\gamma$, so that - again - even small field and density perturbations almost immediately destroy the idealized scenario. In all cases it is the $F_\mathrm{T}$ part of the SG force which must be considered, so that
	\begin{align}
		\Delta \propto 0.05 \left(\frac{m_\mathrm{e}}{m}\right)^2 \lambda_\mathrm{L}[\mu m]^{-1}
	\end{align}
	both for electrons and protons. This distance is in the sub $\mu$m-range for electrons and in the sub pm-range for protons. 
	
	Another interesting point regarding the SG force is, whether all discussed consequences also hold in the limit $\gamma\rightarrow 1$ such that still $|\vb|\approx1$. This case would be important for weakly relativistic protons in laser- and plasma accelerators similar to the recently started project \mbox{JuSPARC}, where polarized proton beams are planned to be created in laser-induced plasma \cite{Huetzen2019, Huetzen2019_a, Buescher2019}. To give a short answer we have a short look at Eqs.(\ref{dtS}) to (\ref{dtV}). Here, for $\gamma=1$, we have to take all terms into account, so that with 
	\begin{align}
		|\Omega_\mathrm{B}|    & \approx 1+a && |\Omega_\mathrm{E}|  \approx a +\frac{1}{2}, && |\Omega_\mathrm{v}| \approx \frac{a}{2},  \\	
		|\Omega_\mathrm{B}'|  &\approx 1     && |\Omega_\mathrm{E}'| \approx \frac{1}{4}, && |\Omega_\mathrm{v}'| \approx \frac{a}{4}, \\
		|\Omega_\mathrm{B}''| &\approx 2     && |\Omega_\mathrm{E}''|\approx \frac{1}{4}, && |\Omega_\mathrm{v}''| \approx \frac{a}{4} \\
	\end{align}	
	we can summarize
	\begin{align}
		&|T_\mathrm{k}| \leq (7 +9a)\partial F, \\ &|E_\mathrm{k}| \leq \frac{5}{4}(4 +a)qF^2, \\ 	&|V_\mathrm{k}| \leq \frac{9}{4}(2 +5a)qF^2.
	\end{align}	
	For $a\ll1$ and $a\approx 1$ these expressions are very close to those in Eqs.(\ref{TE}), (\ref{V}) and definitely in the same order. Thus, to discuss the SG force for weakly relativistic protons, it is sufficient to substitute $\gamma=1$ in the above found scaling laws. 
	
	In summary, we point out that an electron beam has the best chances to be polarized by the relativistic SG force, if the acceleration distance is large enough. For protons we do not see any chance to build up a polarization due to a beam separation - even if pre-factors of $|V_\mathrm{k}|$ ect. are considered. In reverse, this means that the SG force can be neglected in almost all simulations for proton acceleration in lasers and plasma, but not for electrons. It must carefully be estimated whether electrons could interact with a too strong field, where the SG force must be included. If the system passes this check, the temporal evolution of the quasi-classical polarization is completely described by the T-BMT equation.
	
	Originally, the SG force was derived as a quantum mechanical process, but in this section we considered it in a quasi-classical limit. Another quantum-mechanical process, which can lead to self-polarization in storage rings, is the coupling of the spin to the radiation field of the accelerated charges and the thereby caused spin flip. In the next chapter we discuss this effect in a quasi-classical limit again and show whether it must be considered for laser-wakefield accelerators or not.
	
	\section{Sokolov Ternov}\label{ST}
	In storage rings it is possible to observe spontaneous self-polarization of accelerated or stored particle bunches. The mechanism behind it is pure quantum-mechanical and known as the Sokolov-Ternov effect \cite{Sokolov1971}. It describes a polarization build-up due to slightly different probabilities for a spin-flip from down to up $P_\uparrow$ versus spin-flip from up to down $P_\downarrow$ during emission of radiation (Bremsstrahlung). In this chapter we calculate the characteristic polarization time for electrons and protons in laser- and plasma fields. The situation we described is visualized in the third line of Fig.\,\ref{fig:teilchen}, where the particles (red dots) move independently from their individual spin vectors (black arrows) such that an initially unpolarized system state becomes polarized after a certain time. 
	
	To describe the Sokolov-Ternov process, it is necessary to introduce a Lagrangian coupling of the emitted radiation to the spin. The most straight forward approach to do this is to modify Eq.(\ref{Ltot1}) according to
	\begin{align}
		L_\mathrm{total} = L_\mathrm{EM} +L_\mathrm{SG} +L_\mathrm{RAD} +L_\mathrm{ST},
	\end{align}	
	where
	\begin{align}
		L_\mathrm{RAD} = \frac{q}{c}\vb\cdot\Ab_\mathrm{rad} -q\varphi_\mathrm{rad}
	\end{align}	
	describes how the emitted radiation acts back on the particle trajectories and
	\begin{align}
		L_\mathrm{ST} = -\bf{\Omega}_\mathrm{rad}\cdot\sbb
	\end{align}	
	is a direct coupling of spin and radiation via $\bf{\Omega}_\mathrm{rad} = \bf{\Omega}(\Eb_\mathrm{rad}, \Bb_\mathrm{rad})$. To quantize the system and to find the transition probabilities, it is necessary to calculate the corresponding Hamiltonian. It is \cite{Derbenev1973,Jackson1976}
	\begin{align}
		H_\mathrm{total} = H_\mathrm{EM} +H_\mathrm{SG} +H_\mathrm{RAD} +H_\mathrm{ST},
	\end{align}	
	where
	\begin{align}
		&H_\mathrm{EM}  = \gamma mc^2 +q\varphi, && H_\mathrm{SG} = \bf{\Omega}\cdot\sbb,\\
		&H_\mathrm{RAD} = q\varphi_\mathrm{rad} -\frac{q}{c}\vb\cdot\Ab_\mathrm{rad}, && H_\mathrm{ST} = \bf{\Omega}_\mathrm{rad}\cdot\sbb.
	\end{align}	
	
	Now let \ini and \fin be the initial and the final state of a particle that has emitted a soft photon with energy $\hbar\omega\ll\gamma mc^2$. Then we seek to find all matrix elements
	\begin{align}
		M_\mathrm{rad} = \finT H_\mathrm{RAD} \ini, && M_\mathrm{ST}  = \finT H_\mathrm{ST} \ini. \label{M}
	\end{align}	
	In general, $H_\mathrm{ST}$ is regarded as the source of spin-flip radiation. However, since \ini and \fin are not necessarily parallel, also the spin independent term $H_\mathrm{RAD}$ can couple both states in Eq.(\ref{M}), so that both effects must be taken into consideration. Once the probabilities are known, the polarization 
	\begin{align}
		P(t) = P_\mathrm{eq}[1-\exp(-t/\tau_\mathrm{pol})], && P_\mathrm{eq} = \frac{P_\uparrow -P_\downarrow}{P_\uparrow +P_\downarrow} \label{pol}
	\end{align}	
	continuously builds up along the equilibrium-polarization axis $\nb$. The vector $\nb$ is always associated with a given orbital trajectory and is defined to be the explicitly time-independent solution of the T-BMT equation (\ref{PugaSpinNorm}) on that trajectory, so that states quantized along $\nb$ are stationary states. In ordinary storage rings, the polarization time 
	\begin{align}
		\tau_\mathrm{pol}= \frac{1}{P_\uparrow +P_\downarrow}
	\end{align}
	is in the range of minutes to hours, whereas in strong plasma fields, it may be much shorter. The major question concerning us in this work is whether $\tau_\mathrm{pol}$ is small enough compared to the characteristic time scales and whether $P_\mathrm{eq}$ is high enough to achieve a significant polarization.
	
	To estimate the spin-flip probabilities 
	\begin{align}
		P_{\uparrow,\downarrow} \propto \left\langle \int\frac{d\omega}{\hbar\omega}\frac{d\mathcal{P}_{\uparrow,\downarrow}}{d\omega}\right\rangle,
	\end{align}	
	where $\mathrm{d}\mathcal{P}$ is the differential power spectrum of the emitted radiation (for more details see \cite{Schwinger1949, Mane1987}), we have to approximate the averaged transition rates
	\begin{align}
		\alpha_\pm = \frac{1}{4}\int_\mathbb{R}d\tau'\langle0|\overline{[(\omega\cdot\eta)_{t+\tau/2}(\omega\cdot\eta^*)_{t-\tau/2}]_\pm}|0\rangle.
	\end{align}
	These rates are connected to the probabilities via
	\begin{align}
		\alpha_\pm = p_\uparrow \pm p_\downarrow
	\end{align}
	and allow to reformulate the T-BMT equation as
	\begin{align}
		\dot{s}_\mathrm{n} = \hbar^{-1}\alpha_-(s^2-s_\mathrm{n}^2) -\alpha_+s_\mathrm{n}, && s_\mathrm{n} = \sbb\cdot\nb.
	\end{align}
	Here, the diffusion term $\alpha_+s_\mathrm{n}$ is of pure quantum origin and, generally speaking, is a non-local function of the trajectory. An expression for $\alpha_+$ in terms of elementary functions can be obtained only in several limiting situations.
	In the case of practical importance, that of ultra-relativistic motion, when the change of the acceleration is relatively small over the length $\propto|\gamma\dot{\vb}|^{-1}$ in which the radiation is formed, the change of the acceleration is relatively small, and the integral in $\alpha_+$ is concentrated in the region $|\tau|\propto|\gamma\dot{\vb}|^{-1}$ and it can be calculated by using the expansion \cite{Baier1966, Derbenev1973}
	\begin{align}
		\rb(t+\tau) \approx \rb(t) +\vb(t)\tau +\dot{\vb}(t)\frac{\tau^2}{2} +\ddot{\vb}(t)\frac{\tau^3}{6}.
	\end{align}
	
	At $\gamma\gg1$ the radiation, as is well known, is concentrated in a cone with angle $\propto\gamma^{-1}$ about the velocity.
	Taking this circumstance into account and assuming a particle motion across the field lines in a homogeneous field, it is possible to show that \cite{Derbenev1973}
	\begin{align}
		\alpha_- &\approx -\frac{q^2\hbar\gamma^5|\dot{\vb}|^3}{m^2c^8}\left(1+ \frac{14}{3}a +8a^2+\frac{23}{3}a^3 +\frac{10}{3}a^4 +\frac{2}{3}a^5 \right), \\
		\alpha_+ &\approx -\alpha_-\frac{|a|}{a} +R(a), \\
		R &= \frac{q^2\hbar\gamma^5|\dot{\vb}|^3}{e^{\sqrt{12}a}m^2c^8} \left[\left(-1 -\frac{11}{12}a +\frac{17}{12}a^2 +\frac{13}{24}a^3 -a^4\right)\frac{|a|}{a}\right. \nonumber\\
		&\left. +\frac{1}{\sqrt{3}}\left(\frac{15}{8} +\frac{41}{24}a -\frac{115}{48}a^2 -a^3 +\frac{7}{4}a^4\right)\right].
	\end{align}
	These expressions are exact, if the radiation is quasi-classical. Further, for particle with spin $\hbar/2$ the polarization time is $ T_\mathrm{pol} = 1/ \alpha_+(x)$ and the equilibrium polarization is $P_\mathrm{eq} = \alpha_-/\alpha_+$.
	
	If a particle has a small anomalous magnetic moment $a\ll1$, we may neglect all but the lowest order terms, so that 
	\begin{align}
		T_\mathrm{pol}^{-1} &= \lim_{x\rightarrow0} \alpha_+(x) = \frac{q^2\hbar\gamma^5|\dot{\vb}|^3}{m^2c^8} \frac{5\sqrt{3}}{8} \label{small}\\
		P_\mathrm{eq} &= \lim_{x\rightarrow0} \frac{\alpha_-(x)}{\alpha_+(x)} = -\frac{8}{5\sqrt{3}} = -0.92
	\end{align}
	independent of $\gamma$. If the particle has a rather large anomalous magnetic moment $a\gg1$, we only take the last term in $\alpha_-$ into account, so that
	\begin{align}
		T_\mathrm{pol}^{-1} &= T_\mathrm{min}^{-1} = \frac{q^2\hbar\gamma^5|\dot{\vb}|^3}{m^2c^8} \frac{2}{3}a^5 \label{large}\\
		P_\mathrm{eq} &= \lim_{x\rightarrow\infty} \frac{\alpha_-(x)}{\alpha_+(x)} = \frac{q'}{|q'|}
	\end{align}
	which leads to the smallest polarization time possible. In addition to these two cases, which describe the polarization of electron and beams of hypothetical ions with a large core, there exists another important parameter regime, where $|R(a)|\ll |\alpha_-(a)|$. It is the regime of moderate anomalous magnetic momenta, so that $a >1$ is large enough to suppress $R$, but not large enough to simplify the $\alpha_\pm$-functions. The most prominent representative particle for this case is the proton with $a\approx1.8$, so that 
	\begin{align}
		\alpha_-(1.8) &= -128\cdot\frac{e^2\hbar\gamma^5|\dot{\vb}|^3}{m^2c^8} , \\
		R(1.8) &= 0.007\cdot\frac{e^2\hbar\gamma^5|\dot{\vb}|^3}{m^2c^8}
	\end{align}
	and thus
	\begin{align}
		T_\mathrm{pol}^{-1} = \alpha_+(1.8) &\approx 128\cdot\frac{e^2\hbar\gamma^5|\dot{\vb}|^3}{m^2c^8} \ll T_\mathrm{min}^{-1}. \label{proton}
	\end{align}
	Besides these trivial examples, it is worth mentioning that even those particle beams with $\mu=0$ can be spin-polarized due to the spin dependence of radiation. 
	
	Now, it is convenient to derive a scaling law for $T_\mathrm{pol}$ in dependence of the kinetic energy of electron and proton bunches and the strength of the field they are moving in. To find the scaling, we use the same notation as in the previous section, where $F$ was the maximum field strength in the system, and substitute
	\begin{align}
		\frac{|\dot{\vb}|}{c} = \frac{eF}{\gamma mc}
	\end{align}
	for protons and electrons in Eq.(\ref{proton}) and Eq.(\ref{small}) respectively. Then, with $e^2=\alpha\hbar c$, $\hbar c=0.2\text{ GeV\;fm}$ and $\alpha=1/137$, we find
	\begin{align}
		T_\mathrm{pol,proton} = \frac{(m_\mathrm{p}c^2)^7}{128\alpha(\hbar c)^2(eF)^3T_\mathrm{p}^2c} 
	\end{align}
	and
	\begin{align}
		\frac{T_\mathrm{pol,proton}}{T_\mathrm{pol,electron}} = \frac{5\sqrt{3}}{8\cdot128} \frac{\gamma_\mathrm{e}^2}{\gamma_\mathrm{p}^2}\frac{m_\mathrm{p}^5}{m_\mathrm{e}^5}. \label{TeTp}
	\end{align}
	This means that the characteristic polarization time of an initially unpolarized target is $10^{14}$ times larger for protons than for electrons, if both kind of particles have the same $\gamma$ factor. If we compare electrons and protons with equal energy, it is $\gamma_\mathrm{p} m_\mathrm{p}c^2= \gamma_\mathrm{e} m_\mathrm{e}c^2 $ and thus $T_\mathrm{pol,proton} \approx 10^{21} T_\mathrm{pol,electron}$.
	
	The final scaling for protons can be deduced from a simple example, say for high energetic ($T_\mathrm{p} = 100 \text{ GeV}$) protons which are moving in a strong ($F= 10^{17} \text{ V/m}$) field (this field would corresponds to a $10^{24}$\,W/cm$^2$ laser). Because then, with $eF=10^{8}\text{ GeV/m}$, $\gamma_\mathrm{p}=T_\mathrm{p}/(m_\mathrm{p}c^2)$ and $m_\mathrm{p}c^2=1\text{ GeV}$, we find
	\begin{align}
		T_\mathrm{pol,proton} \approx  10^{-5} \text{ s}, \label{TpolP}
	\end{align}
	so that the final scaling law for protons can be formulated as
	\begin{align}
		T_\mathrm{pol,proton} = \frac{10^{14} \text{ s}}{T_\mathrm{p}[\text{GeV}]^2F[\text{TV/m}]^3}.
	\end{align}
	For electrons we find in the same way or with Eq.(\ref{TeTp})
	\begin{align}
		T_\mathrm{pol,electron} = \frac{10^{-7} \text{ s}}{T_\mathrm{e}[\text{GeV}]^2F[\text{TV/m}]^3}.
	\end{align}
	
	If we substitute known numbers from circular accelerators and storage rings, where $F$ is in the order of $10^{-4}$\,TV/m, these two scaling laws predict polarization times in the range of hours for GeV-electrons and more than a million years for TeV-protons. This is the most important reason why electron beams can be spin polarized in storage rings but proton beams cannot. Higher proton energies are available in these days large machines and easily reach the 10 TeV level. However, as long as the field strength is in the sub-TeV regime, the corresponding polarization time remains larger than a million years.
	
	In contrast to circular accelerators and storage rings, plasmas provide thousand times higher field strengths so that the minimum polarization time for GeV-electrons is in the order of microseconds. However, the acceleration distances are in the order of decimeters, corresponding to nano-second particle-field interaction. Hence, the Sokolov-Ternov effect and thus self-polarization can be neglected in laser-plasma-accelerators.
	
	As shown earlier, the minimum depolarization time for proton beams interacting with an ultra-high intense ($10^{24}$\,W/cm$^2$) laser pulse is in the order of microseconds (cmp. Eq.(\ref{TpolP})) but for electrons it is clearly in the sub-fs range. These intensities have been proposed as next-generation ultra-intense lasers and will be available at the Extreme Light Infrastructure \cite{ELI}. In some recent works, which generalize the Sokolov-Ternov effect and discuss polarization effects in the framework of the quantum radiation-reaction regime, it could be shown that electron beams can obtain a notable polarization if they collide with an ultra-intense laser beam \cite{Geng2019, Geng2019a}. Other models, which do not include the Sokolov-Ternov  effect as described in this section but other strong-field QED processes like the strongly nonlinear Compton scattering \cite{Ridgers2014,Piazza2010} and the multi-photon Breit-Wheeler process \cite{Burke1997}, also show that electron beams can be spin polarized if they interact with strong fields from laser-generated QED plasmas \cite{Sorbo2017}. Similar to the Sokolov-Ternov effect, from which we calculated the minimum depolarization time, these models predict polarization times in the fs range and explain that the basic polarization mechanism is an asymmetry in the spin-flip transition rate.
	
	\section{Test Particle Simulations} \label{Test}
	In the last chapters we stated that the polarization of a particle ensemble is always conserved, if all single particle spins precess coherently. In this section we investigate a scenario, where this symmetry is broken and an initial polarization is lost in a certain time. Our basic idea is to measure the depolarization time of an exploding electron cloud radially expanding in a homogeneous electric field. In this setup all electron spins precess individually due to different angles between the electron momenta and the field. Our analytical formulas from section \ref{T-BMT} do not include an angle-specific spin motion, thus we expect larger depolarization times in our simulations than predicted by Eq.\,(\ref{AO5}).
	\begin{figure*}[p]
	\centering
	\subfloat[]{\label{002_ORT_1800}\includegraphics[width=0.45\linewidth]{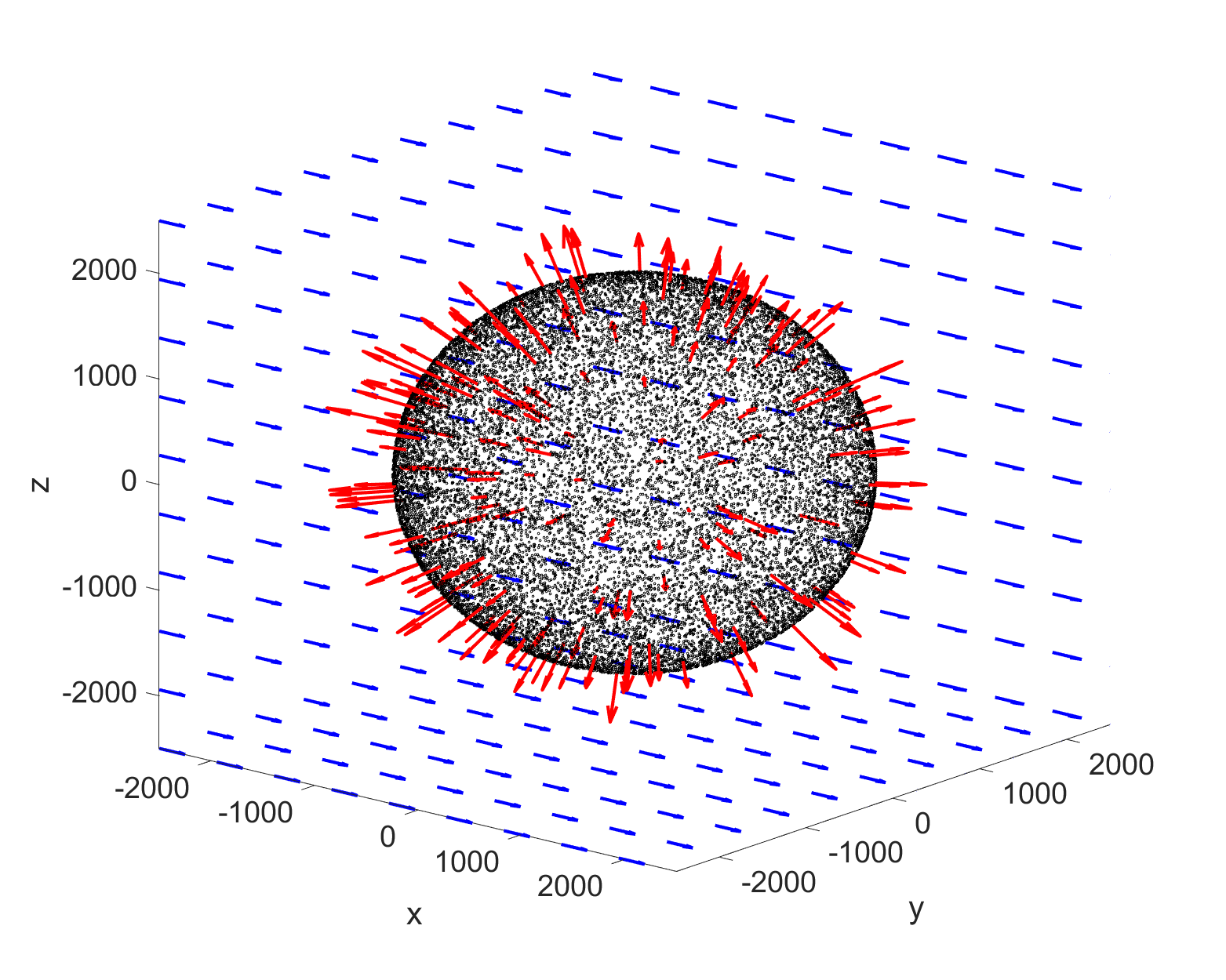}}
	\hfill
	\subfloat[]{\label{012_ORT_1400}\includegraphics[width=0.45\linewidth]{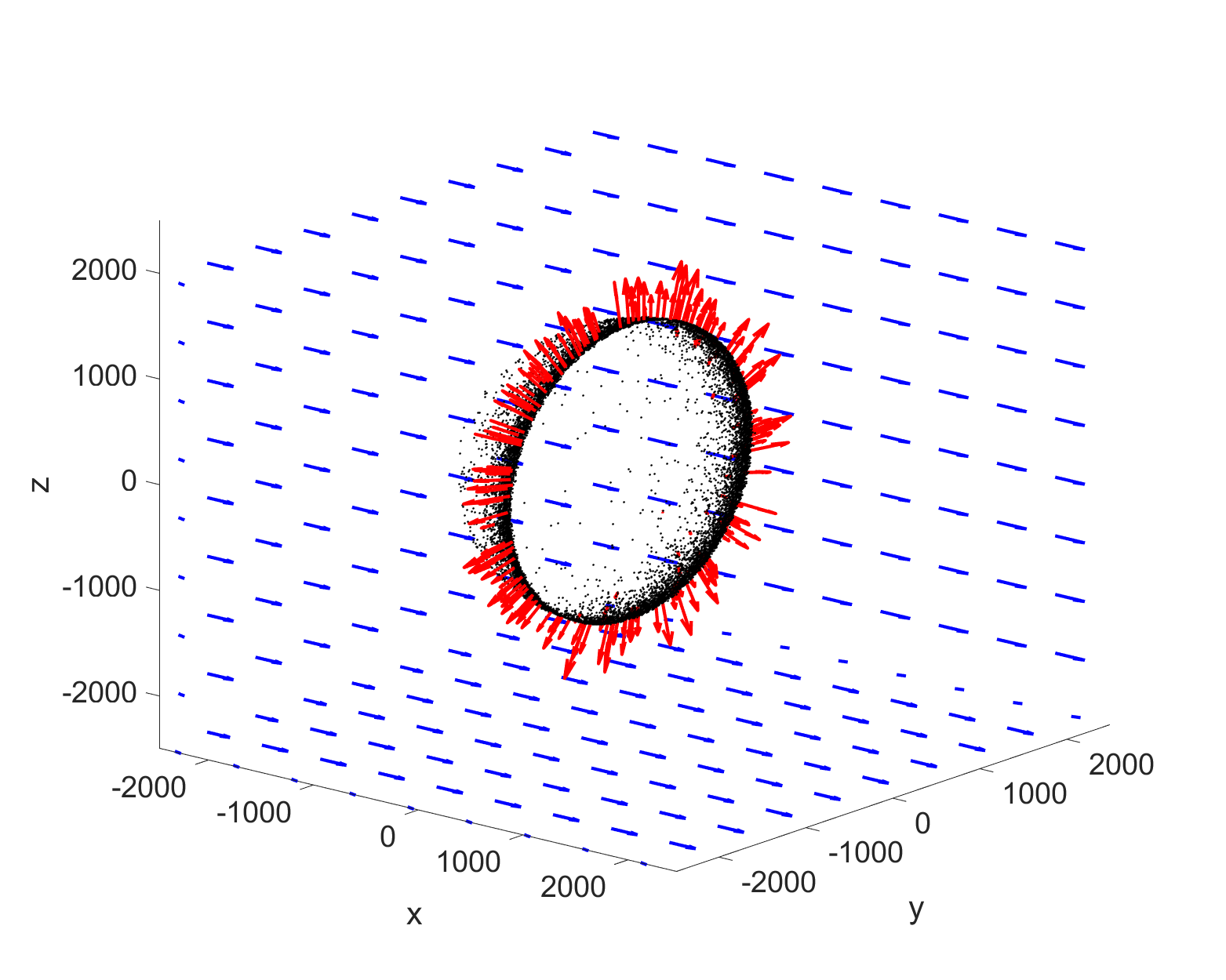}}\\
	\subfloat[]{\label{002_SPIN_500}\includegraphics[width=0.45\linewidth]{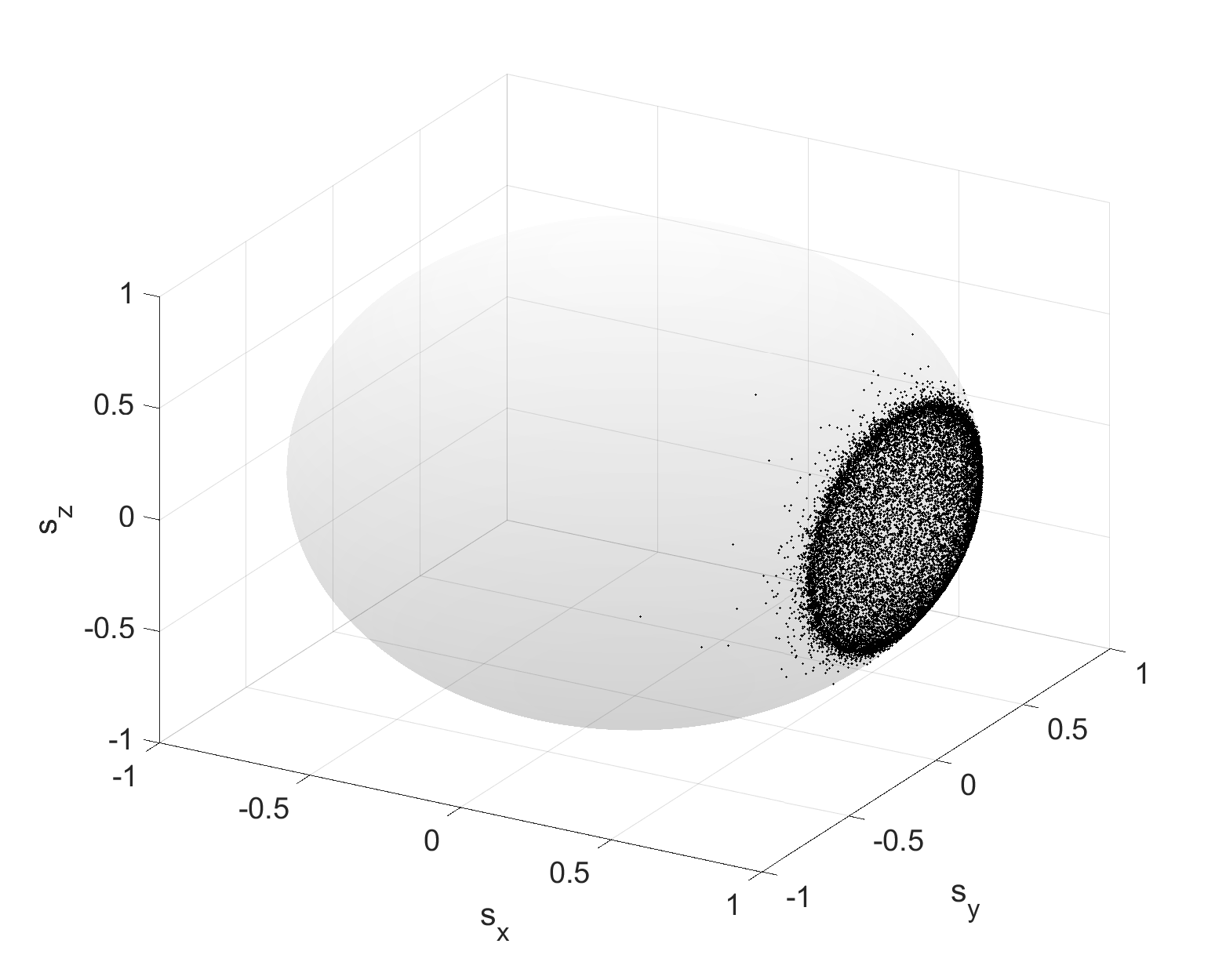}}
	\hfill
	\subfloat[]{\label{012_SPIN_500}\includegraphics[width=0.45\linewidth]{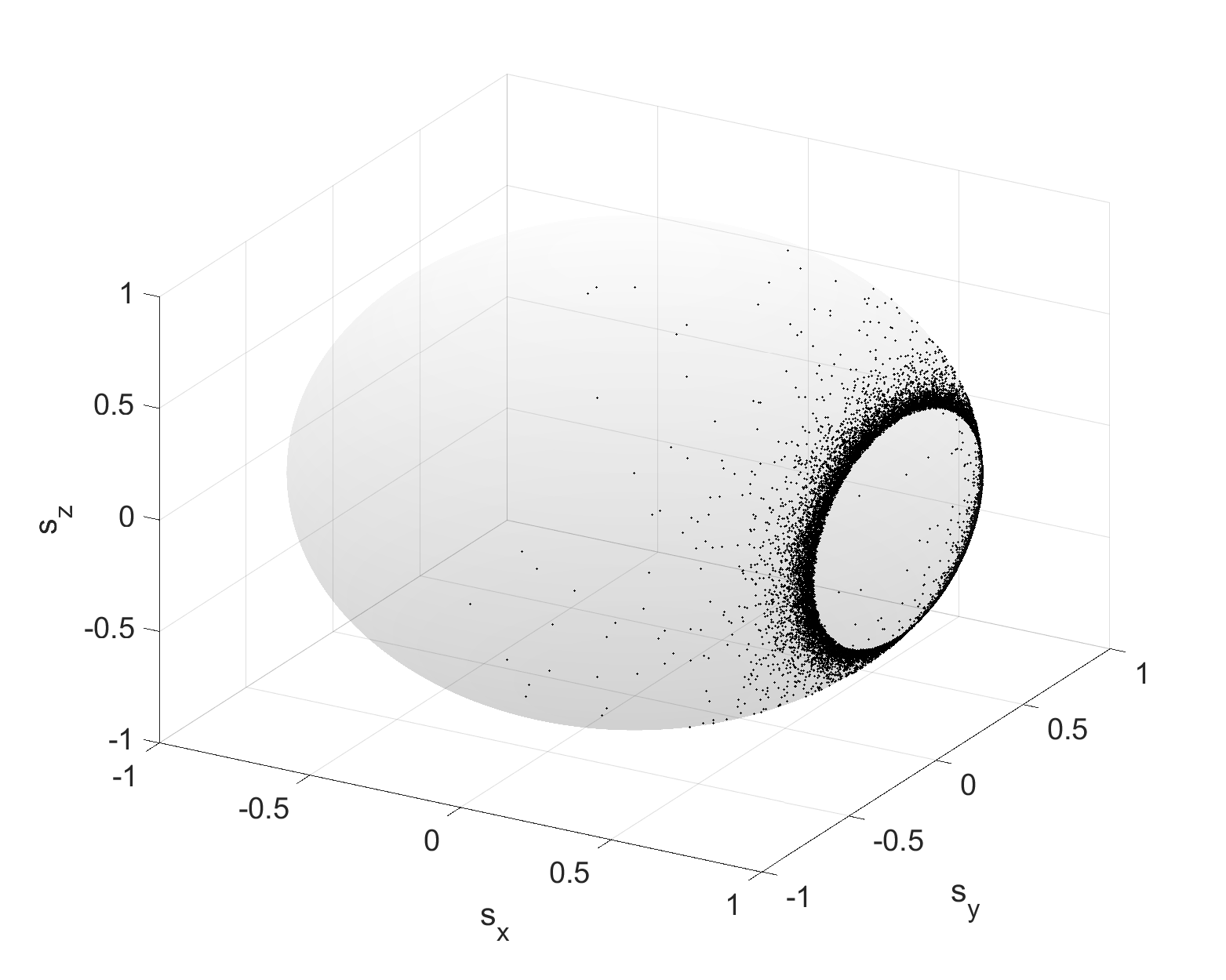}}\\
	\subfloat[]{\label{002_SPIN_1800}\includegraphics[width=0.45\linewidth]{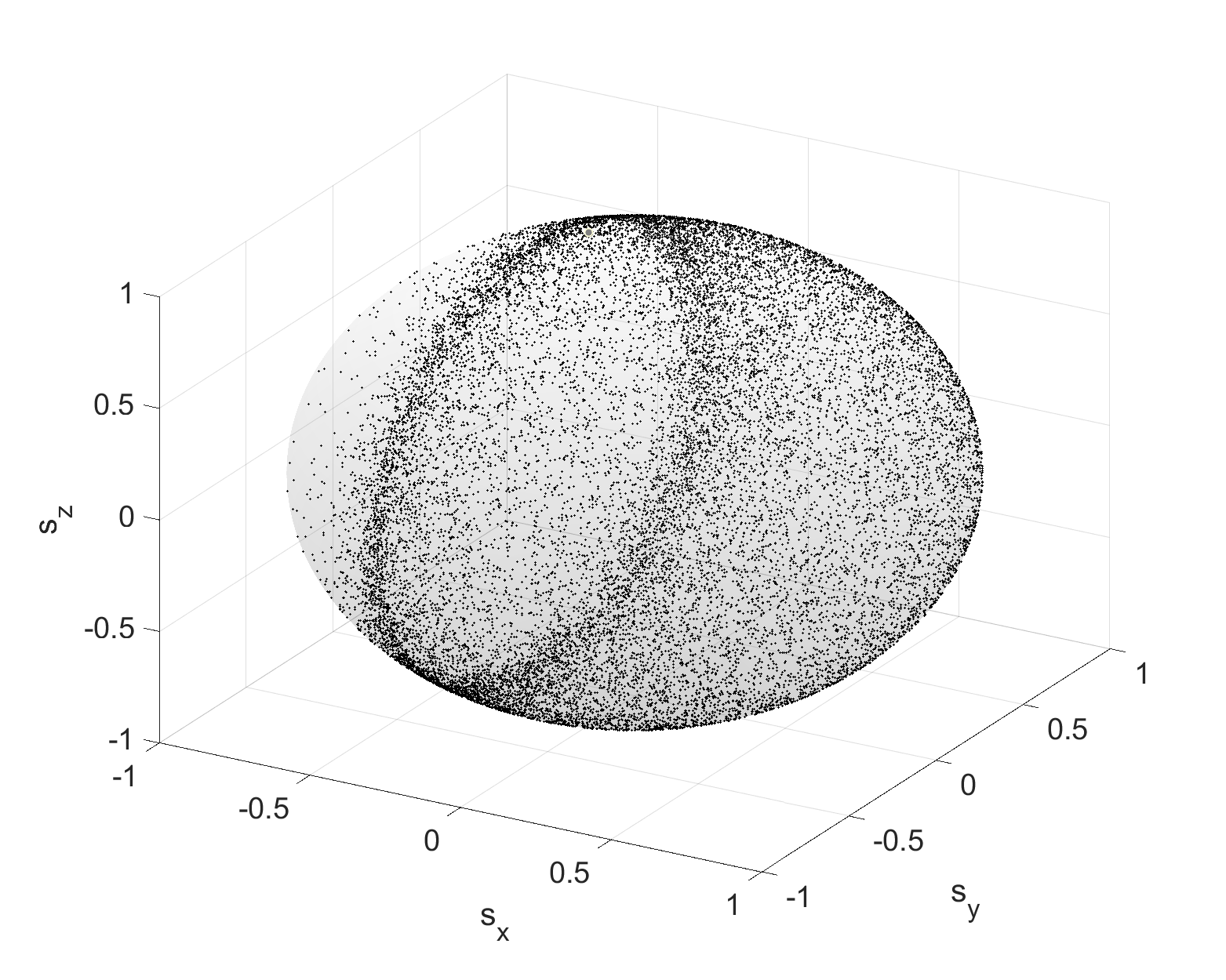}}
	\hfill
	\subfloat[]{\label{012_SPIN_1400}\includegraphics[width=0.45\linewidth]{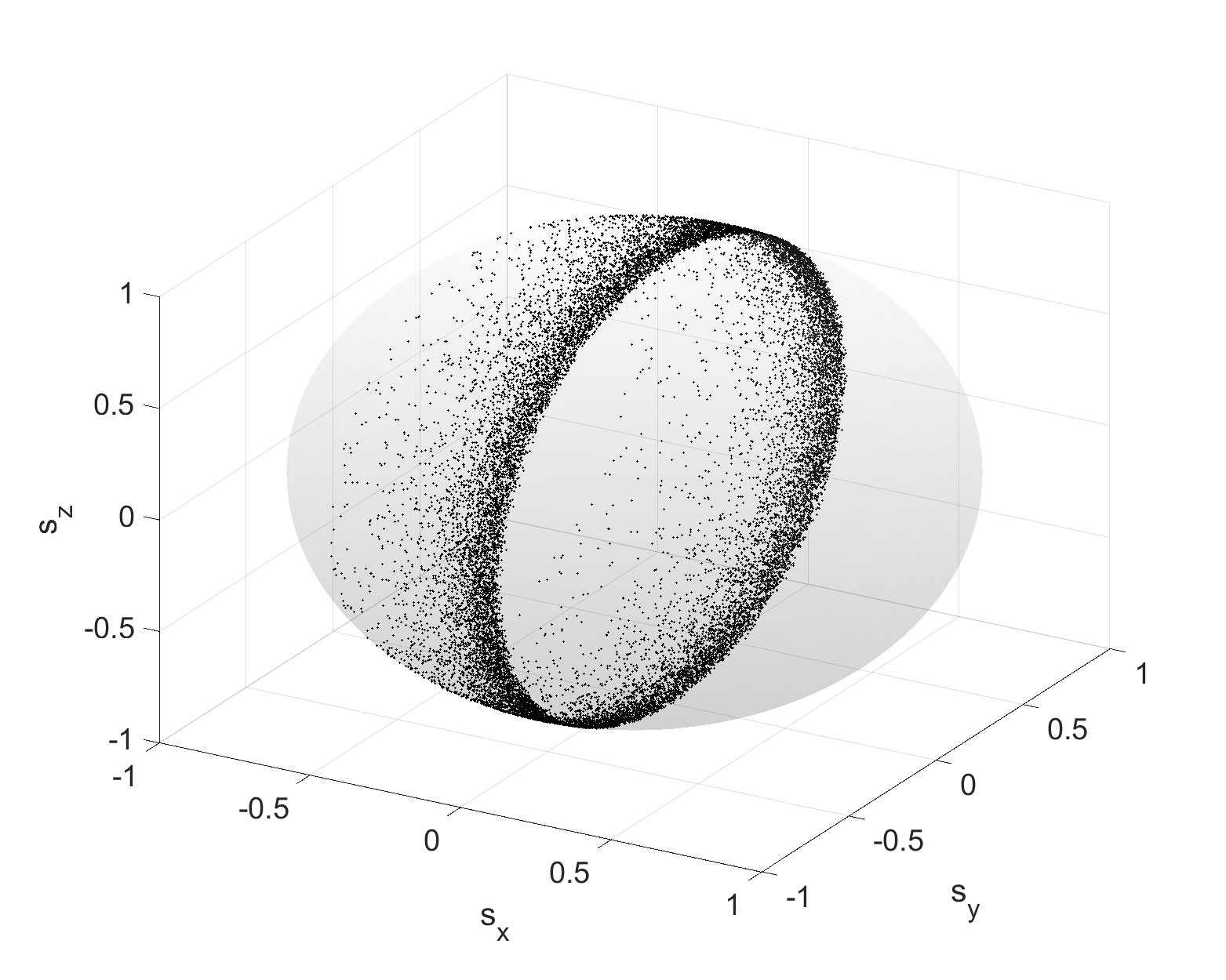}}
	\caption{A high energetic ($\sigma_p=10^4$) electron cloud (black dots) explodes (red arrows indicate moving direction of single electrons) in a homogeneous electric field (blue arrows) while the single particle spins distribute on the unit sphere (gray surface). Left column: The electron cloud expands radially (see red arrows) while the spins create a homogeneous covering of the unit sphere. The simulation time in (\ref{002_ORT_1800}) and (\ref{002_SPIN_1800}) is $t=1800$ which corresponds to the depolarization time (cmp. Fig.\,(\ref{polarization})), while (\ref{002_SPIN_500}) is the spin distribution shortly after the simulation start. Right column: The electron cloud expands cylindrically (see red arrows) in a ring-like structure both in space and in spin space. The simulation time in (\ref{012_ORT_1400}) and (\ref{012_SPIN_1400}) is $t=1400$ which again corresponds to the depolarization time (cmp. Fig.\,(\ref{polarization})), while (\ref{012_SPIN_500}) is the spin distribution shortly after the simulation start.}
	\label{fig:ort_spin}
\end{figure*}

\begin{figure*}[t]
		\centering
		\subfloat[]{\label{polarization}\includegraphics[width=0.49\linewidth]{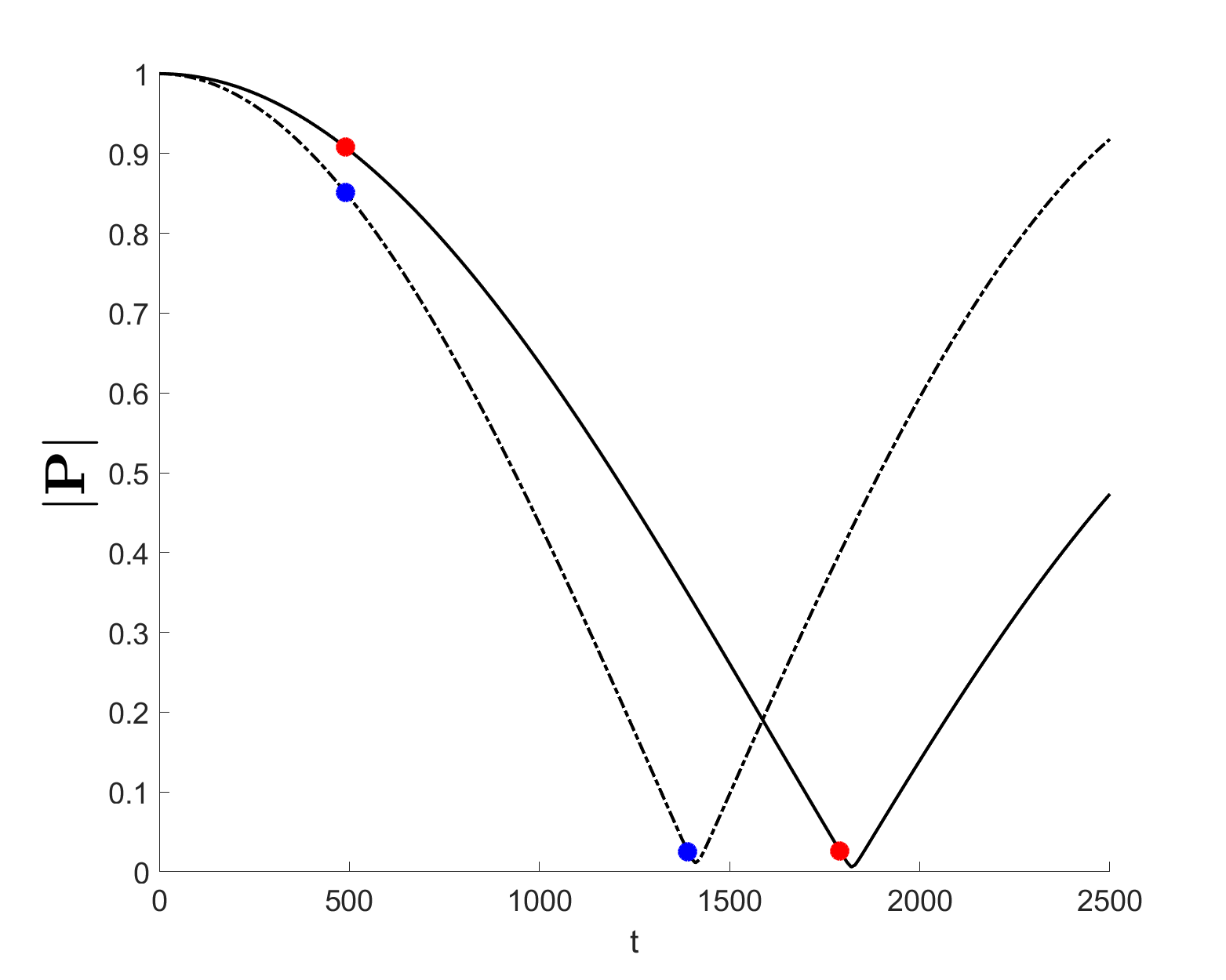}}
		\hfill
		\subfloat[]{\label{TD_s}\includegraphics[width=0.49\linewidth]{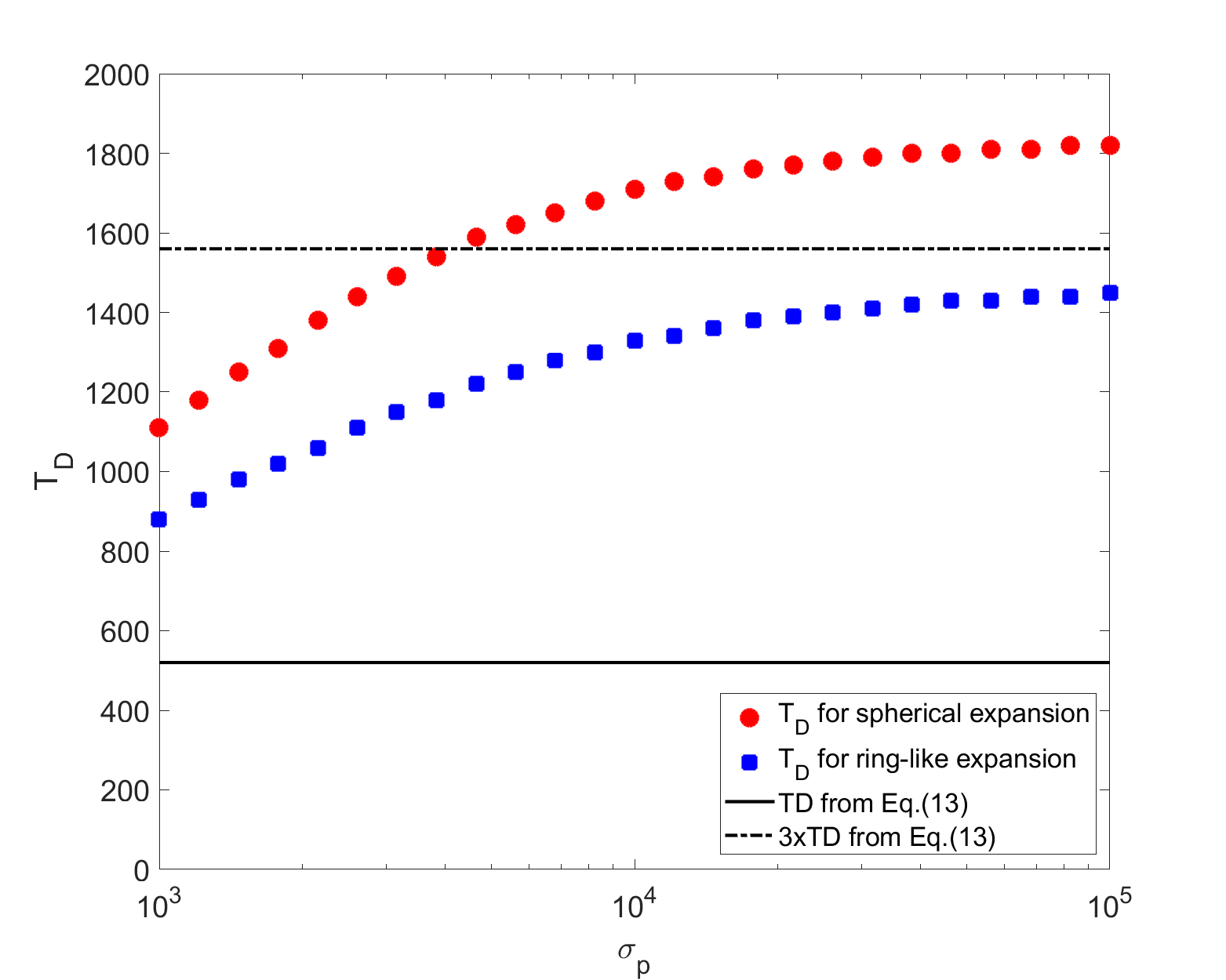}}\\
		\subfloat[]{\label{gamma_mean}\includegraphics[width=0.49\linewidth]{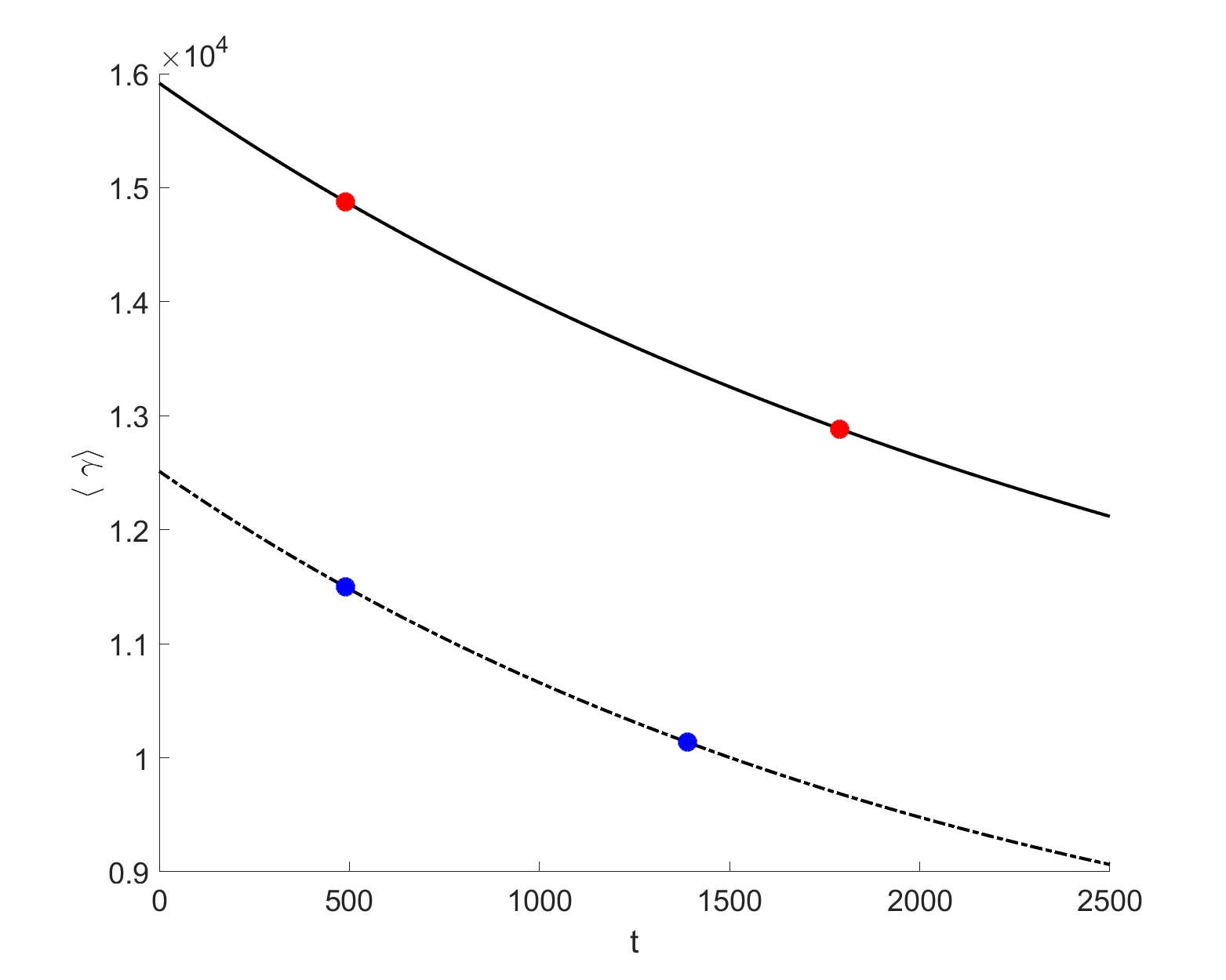}}
		\hfill
		\subfloat[]{\label{CHI_s}\includegraphics[width=0.49\linewidth]{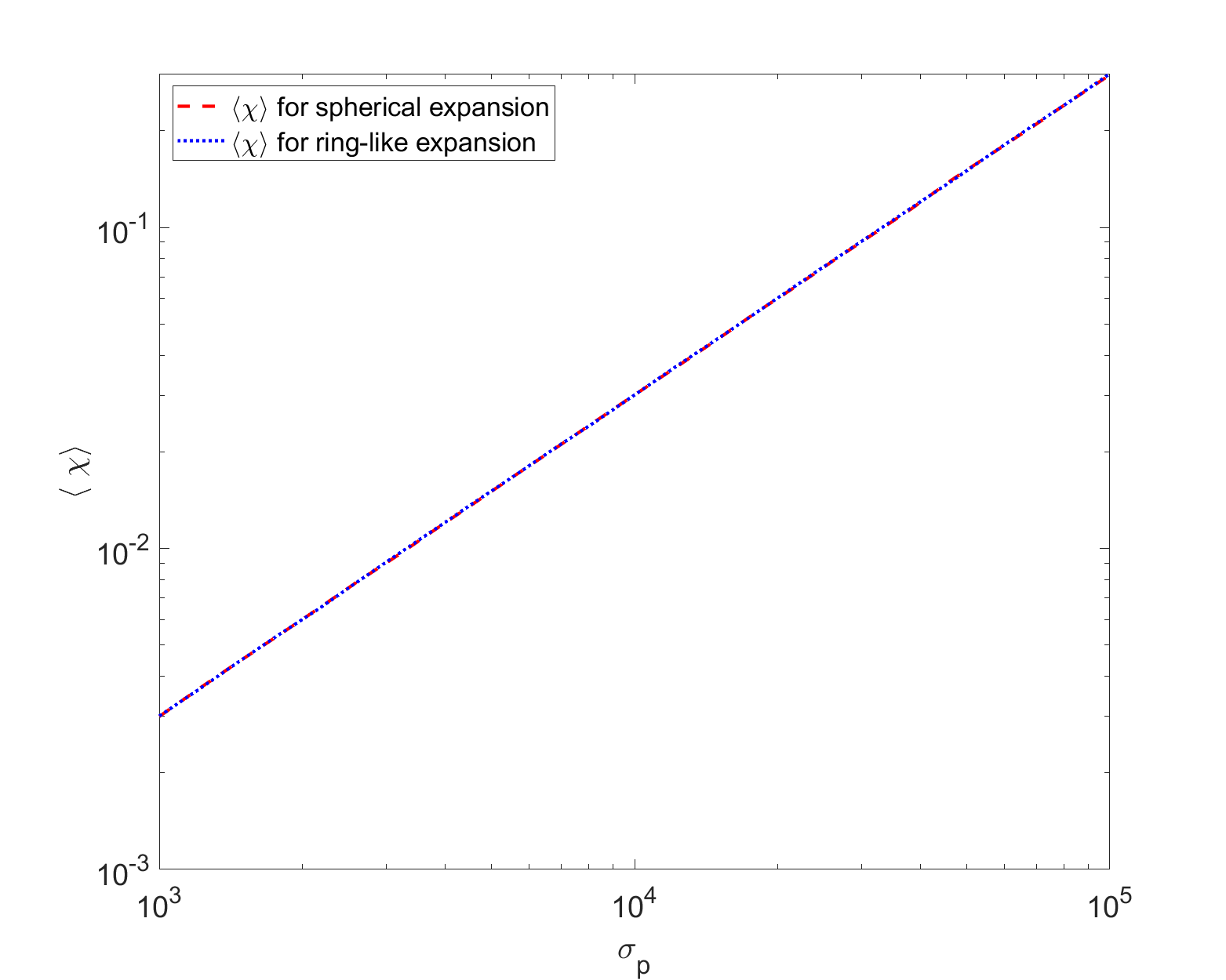}}
		\caption{\ref{polarization}: Temporal evolution of $|\Pb|$ from Eq.(\ref{POL}) of a radially expanding electron cloud (black solid line) and a cylindrically expanding cloud (black dashed line) for $\sigma_p=10^4$. The blue and red time correspond to the spin distributions in Figs.\,(\ref{002_SPIN_500}), (\ref{012_SPIN_500}), (\ref{002_SPIN_1800}) and (\ref{012_SPIN_1400}). \ref{TD_s}: Simulated depolarization time (red and blue dots) compared to the scaling law from Eq.(\ref{AO5}) (black solid line) and a more precise estimation (black dashed line) in dependence of the initial momentum deviation. \ref{gamma_mean}: Temporal evolution of the mean energy in a radially expanding electron cloud (black solid line) and a cylindrically expanding cloud (black dashed line) for $\sigma_p=10^4$. The blue and red time correspond to the spin distributions in Figs.\,(\ref{002_SPIN_500}), (\ref{012_SPIN_500}), (\ref{002_SPIN_1800}) and (\ref{012_SPIN_1400}). \ref{CHI_s}: Dependence of the mean $\chi$-factor on the initial momentum deviation.}
		\label{fig:auswertung}
	\end{figure*}

	To compare our findings from the discussion of the T-BMT equation and the relativistic SG-force to test-particle simulations, we solve the T-BMT equation (\ref{PugaSpinNorm}) for the spin and the equations of motion
	\begin{align}
		\frac{d\rb}{dt} = \frac{\pb}{\gamma}, && \frac{d\pb}{dt} = \Fb_\mathrm{EM} +\Fb_\mathrm{SG} +\Fb_\mathrm{RR} \label{motion}
	\end{align}
	for the electron position and momentum in known external fields. The momentum equation is a combination of the Lorentz force (Eq.\,(\ref{LF})), the relativistic Stern-Gerlach force (Eq.\,(\ref{FSG3})) and a term for the radiation reaction force (RR) taken from Ref.\,\cite{LandauLifschitz}. In a 3D, non-manifest covariant form it can be written as \cite{Tamburini2010}
	\begin{align}
		\Fb_\mathrm{RR} = &-\lambda_\mathrm{RR}\gamma\left[\frac{d\Eb}{dt} +\vb\times\frac{d\Bb}{dt}\right] \nonumber\\ 
		&+\lambda_\mathrm{RR}[(\Eb+\vb\times\Bb)\times\Bb +(\vb\cdot\Eb)\Eb] \nonumber\\
		&-\lambda_\mathrm{RR}\gamma^2[(\Eb+\vb\times\Bb)^2-(\vb\cdot\Eb)^2]\vb.
	\end{align}
	Here
	\begin{align}
		\lambda_\mathrm{RR}=\frac{4\pi}{3}\frac{r_\mathrm{e}}{\lambda}= \frac{2}{3}\frac{e^2\omega}{m_\mathrm{e}c^3}= \frac{4}{3}\alpha_\mathrm{EM}\lambda_\mathrm{SG},
	\end{align}
	$r_\mathrm{e}=e^2/(m_\mathrm{e}c^2)$ is the classical electron radius and $\alpha_\mathrm{EM}=e^2/(\hbar c)$ is the fine-structure constant. The radiation reaction force can be treated classically in our simulations because we choose $\Lambda_\text{SG}=1.2\cdot10^{-6}$. In this situation the ratio between the electric field and the Schwinger field $E_s=m_e^2c^4/(\hbar ce)$ is always small enough to ensure that the $\chi$-parameter
	\begin{align}
		\chi = \frac{\sqrt{(p^\mu F_{\mu\nu})^2}}{m_ecE_s}
	\end{align}	
	is smaller than unity (cmp. Fig.\,(\ref{CHI_s})). The external electric field is pointing in positive \textit{x}-direction and has a normalized constant field amplitude of $E_0=1$. In this way the magnetic field in the particles' rest frame is $\Bb=\gamma|\vb|\textbf{e}_r\times\textbf{e}_x$, which guarantees the desired asynchronous spin motion since each spin sees a different field. Further, we normalize spatial variables to the characteristic wave number $k=eE_0/(m_ec^2)$ and the time to the corresponding frequency $\omega=kc$. 

	In section \ref{T-BMT} we derived a general expression for the depolarization time which ignores all field symmetries and that depends on the maximum field strength of the field-particle system solely. In this chapter the external fields are known so that the maximum spin precession frequency can be calculated more precisely. For example, if we consider a high energetic ($\gamma\gg1/a_e$), initially co-propagating ($\vb||\textbf{e}_\mathrm{z}$) electron, we know the angle between the field and the velocity vector and can estimate that
	\begin{align}
		|\mathbf{\Omega}| &\approx  a_\mathrm{e} |\vb \times \Eb| = 0, \label{Omegamin}
	\end{align}
	such that the expectable minimum depolarization time is arbitrarily large and not comparable to the one in Eq.\,(\ref{AO5}). In contrast to this scenario, where the influence of the field symmetry is dominant, a relativistic electron initially propagating perpendicular to the electric field has $\gamma\gg1/a_e$ but $\vb\bot\textbf{e}_\mathrm{z}$. In this case
	\begin{align}
		|\mathbf{\Omega}| &\approx  a_\mathrm{e} |\vb \times \Eb| \approx a_e, \label{Omegamax}
	\end{align}
	which is of the same order as the estimated spin frequency from Eq.\,(\ref{AO4}) - even if the field symmetry is taken into account. 
		
	In the following we present two simulation series with 20000 initially normal distributed test-electrons with center of mass $\mu_\rb=(0,0,0)^t$, mean momentum $\mu_\pb=(0,0,0)^t$, standard deviation $\sigma_\rb=(1,1,1)^t$ and identical spin $\sbb=\text{e}_\mathrm{x}$. In the first simulation series we consider an exploding electron cloud with momentum deviation  $\sigma_\pb=\sigma_p(1,1,1)^t$ which expands spherically. In the second series we change the momentum deviation to $\sigma_\pb=\sigma_p(0,1,1)^t$ such that all electrons move initially perpendicular to the electric field. To solve Eqs.\,(\ref{PugaSpinNorm}) and (\ref{motion}) we adopt the Boris push operator splitting method \cite{Boris1970} in the following way: \\\\
	1. First half position update:
	\begin{align}
		\rb^{n+1/2} = \rb^n +\frac{\pb^n}{2\gamma^n}\Delta t
	\end{align}
	2. First half momentum acceleration:
	\begin{align}
		\pb^{-} = \pb^n +\frac{\Fb_\mathrm{acc}(\rb^{n+1/2},\pb^n,\sbb^n)}{2}\Delta t
	\end{align}
	3. Momentum rotation step:
	\begin{align}
		\pb^{+} = \pb^{-} +(\pb^{-}+(\pb^{-}\times\ab_\mathrm{p}))\times\tb_\mathrm{p}
	\end{align}
	4. Spin rotation step:
	\begin{align}
		\sbb^{n+1} = \sbb^n +(\sbb^n+(\sbb^n\times\ab_\mathrm{s}))\times\tb_\mathrm{s}
	\end{align}
	5. Second half momentum acceleration:
	\begin{align}
		\pb^{n+1} = \pb^{+} +\frac{\Fb_\mathrm{acc}(\rb^{n+1/2},\pb^{+},\sbb^{n+1})}{2}\Delta t
	\end{align}
	6. Second half position update:
	\begin{align}
		\rb^{n+1} = \rb^{n+1/2} +\frac{\pb^{n+1}}{2\gamma^{n+1}}\Delta t
	\end{align}
	For the momentum rotation we use the instantaneous rotation axis
	\begin{align}
		\ab_\mathrm{p} = \left[\frac{\Bb^{n+1/2}}{\gamma^{-}} +\lambda_\mathrm{RR}\frac{d\Bb^{n+1/2}}{dt}\right]\frac{\Delta t}{2}
	\end{align}
	and $\tb_\mathrm{p}=2\ab_\mathrm{p}/(1+|\ab|_\mathrm{p}^2)$, while the spin is rotated with $\ab_\mathrm{s}=-\mathbf{\Omega}$ and $\tb_\mathrm{s}=2\ab_\mathrm{s}/(1+|\ab\mathrm{s}|^2)$ likewise. The vector $\Fb_\mathrm{acc}$ is the sum of all those force terms which do not rotate the momentum vector, i.e.
	\begin{align}
		\Fb_\mathrm{acc}(\rb,\pb,\sbb) = \frac{d\pb}{dt} -\frac{2\pb\times\ab_\mathrm{p}}{\Delta t}.
	\end{align}
		
	An exemplary evolution of the spatial electron distribution and the spin distribution for $\sigma_p=10^4$ is shown in Fig.\,(\ref{fig:ort_spin}). The first picture in the left column shows the electron distribution of the radially expanding cloud of the first simulation series at time $t=1800$, which is close to the actual depolarization time (cmp. Fig.\,(\ref{polarization})). Due to the geometry of the system, all possible angles between single particle velocities and the electric field can be found which, according to Eqs.(\ref{Omegamin}) and (\ref{Omegamax}), leads to a distribution of the spin frequencies between $\Omega_\text{min}=0$ and $\Omega_\text{max}=1/a_e$. As can be seen in Figs.\,(\ref{002_SPIN_500}) and (\ref{002_SPIN_1800}), the distribution is similar to a homogeneous coating covering the unit sphere including a dense ring-like layer at its left hand side. The temporal evolution of the polarization of this system is shown as black solid line in Fig.\,(\ref{polarization}). The first red dot indicates the polarization at time $t=500$ for the spin distribution in Fig.\,(\ref{002_SPIN_500}) shortly after the simulation start. The second red dot marks the polarization at time $t=1800$ for the coating in Fig.\,(\ref{002_SPIN_1800}), which is close the actual depolarization time, but more than three times larger than the predicted one from the scaling law in Eq.(\ref{AO5}) (also cmp. Fig.\,(\ref{TD_s})). A more precise estimation of the depolarization time would be $T_\text{D}\approx3\pi/(6.6a_\text{e}F)$, because the vanishing  magnetic field reduces the number of potentially non-zero terms in the T-BMT equation by a factor of three. However, this time is still smaller than the observed one because the dense ring-like electron layer must compensate the homogeneous coating of slowly varying spins.
	
	A similar observation can be formulated for the second simulation series for electrons which move initially perpendicular to the electric field. As seen in the first picture in right column of Fig.\,(\ref{fig:ort_spin}), the spatial distribution is a dense ring-like structure with a thin electron cloud gradually accelerating in negative $x$-direction. At the beginning of the simulation, all spins precess with the same frequency due to their initially perpendicular velocities. This leads to a spin distribution resembling a dense ring-like structure without any coating covering the unit sphere in Fig.\,(\ref{012_SPIN_500}). At later times this structure dissolves since some single spins precess slightly faster than the major part of the system (cmp. Fig.\,(\ref{012_SPIN_1400})). The temporal evolution of the polarization of this system is shown as black dashed line in Fig.\,(\ref{polarization}). Here, the first blue dot indicates the polarization at time $t=500$ for the spin distribution in Fig.\,(\ref{012_SPIN_500}) shortly after the simulation start. The second blue dot marks the polarization at time $t=1400$ for the dissolving ring in Fig.\,(\ref{012_SPIN_1400}), which is close the actual depolarization time. In contrast to the first simulation, this time is less than $T_\text{D}\approx3\pi/(6.6a_\text{e}F)$ but still larger than the predicted time from Eq.(\ref{AO5}) (also cmp. Fig.\,(\ref{TD_s})).
	
	If we vary the initial momentum standard deviation in the first simulation series, we observe a similar polarization evolution as in Fig.\,(\ref{polarization}) for $\sigma_p > 5\cdot10^3$. For smaller standard deviations, the estimation $T_\text{D}\approx3\pi/(6.6a_\text{e}F)$ is too high because the condition $\langle\gamma\rangle\ll1/a_\text{e}$ for Eq.(\ref{AO5}) is no longer fulfilled (cmp. red dots in Fig.\,(\ref{TD_s})). In contrast to that, the observed depolarization times in second simulation series (see blue dots in Fig.\,(\ref{TD_s})) are always less than $3\pi/(6.6a_\text{e}F)$. However, the scaling law $\min T_\text{D}=\pi/(6.6a_\text{e}F)$ is valid for all simulations in both series. If the initial mean electron energy is high enough, both systems are depolarized before the gamma factor becomes small due to radiation effects. If the gamma factor is rather low but still high enough to meet the condition $\langle\gamma\rangle\ll1/a_\text{e}$ at the beginning of the simulation, the energy loss due to radiation is limited (see Fig.\,(\ref{gamma_mean})) and the condition is met till depolarization.

	\section{Conclusion}
	In this work we derive scaling laws for the (de-)polarization time of high energetic particle beams moving in strong fields. We discuss all relevant mechanisms, namely an asynchronous spin precession, the Sokolov-Ternov effect and the Stern-Gerlach force, that may have an influence on the polarization. We consider field strengths which are comparable to those known from present days laser and plasma accelerators for light electrons and comparatively heavy protons. In our derivations we use rather general approximations of the field gradient and the field strength so that our results are independent from a specific field configuration and the specific particle moving direction. 
	
	Our scaling law for the minimum depolarization time for initially polarized electron and proton beams is calculated from the T-BMT equation under the assumption that all particle spins precess incoherently. A comparison to test-particle simulations of high energetic electrons moving in a homogeneous electric field shows that the observed depolarization time scales as predicted. In any case, the scaling law must always be interpreted as a lower limit.
	
	A discussion of the generalized Stern-Gerlach force shows that the single-particle trajectories are perturbed rather by a T-BMT-rotation-induced spin change than due to a coupling of the spin to the energy- or velocity-changing rates, while even small field variations must be taken into account. Regarding a possible polarization build up through spin-depending beam split up effects, we see that a TeV electron beam has the best chances to be polarized when the plasma is dense enough and the acceleration distance (time) is large enough. For protons we do not see any chance to build up a polarization by a beam separation. 
	
	After having discussed the Sokolov-Ternov effect in the framework of a Hamiltonian theory, we formulate scaling laws for the minimum polarization time for electrons and protons in arbitrary fields. Applied to conventional accelerators we recover known polarization times, while our scalings predict that electrons moving in strong ($\approx 10^{17}$\,V/m) fields should be polarized in less than a fs. Theoretical models incorporating strong-field QED effects agree with our findings.
	
	In the last chapter of our work we present a couple of test-particle simulations to investigate a scenario, where an initial polarization of a relativistic electron beam is lost in a certain time. In our simulations an electron cloud expands in a homogeneous electric field such that incoherent spin oscillations can be guaranteed. As mentioned above, the observed depolarization time scales as predicted by the scaling law we derived from the T-BMT equation.

	\section{Acknowledgement}
	This work has been carried out in the framework of the \textit{Ju}SPARC (J\"ulich Short-Pulse Particle and Radiation Center) project and has been supported by the ATHENA (Accelerator Technology HElmholtz iNfrAstructure) consortium. The Chinese authors acknowledge support through the Strategic Priority Research Program of Chinese Academy of Sciences (Grant No. XDB 16010000), the National Science Foundation of China (No. 11875307) and the Recruitment Program for Young Professionals. 
	
	\appendix
	\section{Vector identities for the derivation of the Stern-Gerlach force}\label{APP_SG}
	In section \ref{SG} we derived the relativistic generalization of the Stern-Gerlach force. In this appendix we summarize all needed vector identities.
	\\
	
	Let $\ab$ and $\bb$ be two arbitrary non-zero constant vectors. Then
	\begin{align}
		\nabla_\vb [(\vb\cdot\ab)(\vb\cdot\bb)] = \ab(\vb\cdot\bb) +(\vb\cdot\ab)\bb
	\end{align}
	and
	\begin{align}
		\nabla_\vb [(\vb\times\ab)\cdot\bb] = \ab\times\bb. \label{vrot}
	\end{align}
	Let $\gamma=1/\sqrt{1-|\vb|^2}$ be the Lorentz factor. Then $\vb$ is the velocity, $\pb=\gamma\vb$ is the kinetic momentum and \begin{align}
		\nabla_\vb \frac{1}{\gamma} = -\pb, && \nabla_\vb \gamma = -\gamma^2\nabla_\vb\frac{1}{\gamma} = \gamma^2\pb.
	\end{align}
	Further, it is
	\begin{align}
		\nabla_\vb \frac{1}{1 +\gamma} = -\frac{\gamma^2}{(1+\gamma)^2}\pb
	\end{align}
	and 
	\begin{align}
		\nabla_\vb\frac{\gamma}{1+\gamma} = \frac{\gamma	^2}{(1+\gamma)^2}\pb.
	\end{align}
	Let $\Fb(\rb)$ be a vector-valued function and $\ab$,$\bb$ two arbitrary non-zero vectors. Then 
	\begin{align}
		|\nabla(\Fb\cdot\ab)| & \leq |\ab|\sqrt{\sum_{i=1}^3 (\partial_{x_i}|\Fb|)^2 } \nonumber\\
		& \leq |\ab|\sqrt{\sum_{i=1}^3 \left(\sum_{j=1}^3\frac{|F_j|}{|\Fb|}|\partial_{x_i}F_j|\right)^2 } \nonumber\\
		& \leq 3\sqrt{3}|\ab|\partial F, \label{grad_Fa}
	\end{align}
	where
	\begin{align}
		\partial F = \max_{i,j=1,2,3}(|\partial_{x_i}F_j|)	. \label{dF}
	\end{align}
	The generalization of Eq.(\ref{vrot}) for an arbitrary vector-valued function can be formulated as
	\begin{align}
		\nabla[(\Fb\times\ab)\cdot\bb] & = \sum_{i=1}^3 \left(\frac{\partial\Fb}{\partial x_i}\cdot(\ab\times\bb) \right)\hat{e}_{x_i},
	\end{align}
	so that Eq.(\ref{dF}) gives
	\begin{align}
		|\nabla[(\Fb\times\ab)\cdot\bb]| & \leq \sum_{i=1}^3 \left(\sqrt{3}\partial F |\ab||\bb| \right)^2\hat{e}_{x_i} \\
		& = 3|\ab||\bb|\partial F.
	\end{align}
	Some important transformations that appear throughout the MS are
	\begin{align}
		\frac{d}{dt} & = \frac{\partial}{\partial t} + \vb\cdot\nabla +\dot{\vb}\cdot\nabla_\vb, \label{dt}\\
		\dot{\vb}\cdot\nabla_\vb & = \sum_{i=1}^3\dot{v}_i\frac{\partial\gamma}{\partial v_i}\frac{\partial}{\partial\gamma} = \dot{\gamma}\frac{\partial}{\partial\gamma} \\
		\dot{\gamma} & = \frac{\pb\cdot\dot{\pb}}{\gamma} = q\vb\cdot\Eb  \\
		\dot{\vb} & = \frac{\dot{\pb}}{\gamma} -\frac{\pb}{\gamma^2}\dot{\gamma} = \frac{q}{\gamma}[\Eb +\vb\times\Bb -(\vb\cdot\Eb)\vb ] \label{dv} 
	\end{align}
	
	For derivation of $V_k$ we need the following identities
	\begin{align}
		|\dot{\vb}\cdot\Bb| & \leq \frac{q}{\gamma}(1 +|\vb|^2)|\Eb||\Bb| < \frac{q}{\gamma}2F^2\\
		|\dot{\vb}\times\Eb| & \leq \frac{q}{\gamma}[|\vb||\Eb||\Bb| +|\vb|^2|\Eb|^2] < \frac{q}{\gamma}2F^2 \\
		|\dot{\vb}| & \leq \frac{q}{\gamma}[|\Eb| +|\vb||\Bb| +|\vb|^2|\Eb|] < \frac{q}{\gamma}3F \label{vdot}
	\end{align}

\end{document}